\documentclass[12pt,a4paper]{article}
\usepackage{url}
\usepackage[dvipdfmx]{graphicx}
\usepackage{amsmath,amssymb}
\usepackage{mathrsfs}
\usepackage{comment}
\usepackage{cancel}
\usepackage{xcolor}
\usepackage{cite}
\usepackage{arydshln}
\usepackage[italicdiff]{physics}
\usepackage[colorlinks=true,citecolor=green!60!blue,urlcolor=black]{hyperref}
\renewcommand\thefootnote{\#\arabic{footnote}}
\numberwithin{equation}{section}
\begin{document}

\begin{titlepage}

\def\thefootnote{\fnsymbol{footnote}}

\begin{center}

\hfill KANAZAWA-22-05 \\
\hfill November, 2022

\vspace{0.5cm}
       {\Large\bf Subcritical regime of hybrid inflation\\with modular $A_4$ symmetry }

\vspace{1cm}
{\large Yoshihiro Gunji,}$^{\it (a)}$ 
{\large Koji Ishiwata,}$^{\it (a)}$
{\large Takahiro Yoshida}$^{\it (b),\,(c)}$

\vspace{1cm}

{\it $^{(a)}$Institute for Theoretical Physics, Kanazawa University, Kanazawa
  920-1192, Japan}

{\it $^{(b)}$Department of Information, Kaishi Professional University, Niigata 950-0916, Japan}

{\it $^{(c)}$Department of Physics, Niigata University, Niigata 950-2181, Japan}

\vspace{1cm}

\abstract{We consider a supergravity model that has the modular $A_4$
  symmetry and discuss the interplay between the neutrino mixing and
  inflation. The model contains right-handed neutrinos that have the
  Majorana masses and additional Yukawa couplings to the waterfall
  field.  In the model an active neutrino is massless and we find that
  only the inverted hierarchy is allowed and the Majorana phase is
  predicted to be around $\pm (120\text{--}180)^\circ$ from the
  observed neutrino mixing data.  In the early universe, one of
  right-handed sneutrinos plays the role of the inflaton
  field. Focusing on the subcritical regime of the hybrid inflation
  that is consistent with the cosmic microwave background data, we
  analyze the dynamics of the scalar sector and derive an upper bound
  $\order{10^{10}}~{\rm GeV}$ on the scale of the Majorana mass.  }

\end{center}
\end{titlepage}

\section{Introduction}

Inflation, a hypothesis of accelerated expansion of the Universe, is
supported by the observation of the cosmic microwave background (CMB)
radiation. It is driven by the vacuum energy at the early state of the
Universe and it is realized by a slowly-rolling scalar field, called
inflaton.  According to decades of the observations and analysis of
the CMB, some properties of inflation have been revealed. The latest
results by Planck collaboration~\cite{Akrami:2018odb} report that the
amplitude $A_s$ and spectral index $n_s$ of the scalar perturbation
and the ratio $r$ of the tensor mode to the scalar mode are given by
 $ A_s = (2.0989\pm 0.029)\times 10^{-9}~~ (68\%\,{\rm CL})$, 
 $ n_s = 0.9649\pm 0.0042~~ (68\%\,{\rm CL})$,  and
  $r <0.056~~(95\%\,{\rm CL})$.
The results already exclude some of inflation models and future
experiments are expected to measure the observables with more
precision.

Besides the accelerated expansion of the Universe, the inflaton field
plays another important role, i.e., reheating to create the radiation
dominated Universe. Then the production of dark matter and baryons
follows. Thus, it is tempting to consider an inflation model that
provide the sequence of the thermal history after
inflation. Moreover, the model would be motivated if
it is controlled by an underlying symmetry.  The modular symmetry and
supersymmetry are the promising candidates.

The modular symmetry is the discrete symmetry associated with the
compactification of the extra dimensions inspired by superstring
theory. It has recently caught attention from the phenomenological
point of view since Ref.~\cite{Feruglio:2017spp} pointed out that it
gives a consistent pattern of the neutrino mixing
data~\cite{Adamson:2013whj,Adamson:2013ue,Abe:2017vif,Abe:2018wpn,Adamson:2017gxd,NOvA:2018gge}. Variety types of symmetries have been studied so far; for instance,
the modular $S_3$~\cite{Kobayashi:2018vbk},
$A_4$~\cite{Feruglio:2017spp, Criado:2018thu, Kobayashi:2018scp,
  Kobayashi:2018wkl, Novichkov:2018yse, Okada:2019uoy,
  Kobayashi:2019mna, Ding:2019zxk, Okada:2019mjf, Kobayashi:2019xvz,
  Chen:2019ewa, Zhang:2019ngf, Nomura:2019xsb, Kobayashi:2019gtp,
  Wang:2019xbo, King:2020qaj, Abbas:2020qzc, Okada:2020rjb,
  Nomura:2020opk, Nomura:2020cog, Asaka:2020tmo, Nagao:2020snm,
  Okada:2020brs, Gehrlein:2020jnr, Hutauruk:2020xtk, Yao:2020qyy,
  Feruglio:2021dte, Kobayashi:2021jqu, Tanimoto:2021ehw,
  Nomura:2021yjb, deMedeirosVarzielas:2021pug, Chen:2021prl,
  Okada:2021aoi, Nomura:2021pld, Liu:2021gwa, Kikuchi:2022txy,
  Nomura:2022hxs, Otsuka:2022rak, Kobayashi:2022jvy, Ahn:2022ufs,
  Kashav:2022kpk, Nomura:2022boj, Ishiguro:2022pde, Gogoi:2022jwf},
$A_5$~\cite{Novichkov:2018nkm, Ding:2019xna,
  deMedeirosVarzielas:2022ihu}, and other modular
groups~\cite{Nilles:2020nnc, Ding:2020msi, Li:2021buv, Ohki:2020bpo}.
Quark masses and mixings are investigated in Refs.\,
\cite{Okada:2018yrn, Kuranaga:2021ujd, Kikuchi:2022geu, Kikuchi:2022svo}. In the
modular symmetric framework, soft supersymmetry breaking
terms~\cite{Kikuchi:2022pkd} and modular
stabilization~\cite{Kobayashi:2019uyt, deMedeirosVarzielas:2020kji,
  Ding:2020zxw, Ishiguro:2020tmo, Novichkov:2022wvg} are also
discussed. For other phenomenological applications, for example, the
application of modular symmetry to grand-unified theories is
discussed in Refs.~\cite{deAnda:2018ecu, Kobayashi:2019rzp,
  Du:2020ylx, Zhao:2021jxg, Chen:2021zty, King:2021fhl, Ding:2021zbg,
  Ding:2021eva, Kobayashi:2022sov}. Moreover, in cosmology, models with dark matter
candidates~\cite{Nomura:2019jxj, Nomura:2019yft, Okada:2019xqk,
  Nomura:2019lnr, Okada:2019lzv, Okada:2020dmb, Nagao:2021rio,
  Kobayashi:2021ajl} have been considered and Refs.~\cite{Asaka:2019vev,
  Behera:2020sfe, Mishra:2020gxg, Kashav:2021zir, Okada:2021qdf,
  Dasgupta:2021ggp, Behera:2022wco, Kang:2022psa, Ding:2022bzs} have applied it to leptogenesis.

In the present study we apply the modular $A_4$ symmetry to inflation.
Introducing three right-handed (s)neutrinos, we consider one of
right-handed sneutrinos plays the role of the inflaton field. Since
the typical model of the right-handed sneutrino inflation tends to be
chaotic inflation due to the quadratic term, which is already excluded
by the CMB observations, we extend the model to the $D$-term hybrid
inflation by introducing a new Yukawa term.  These days variety of new
$D$-term hybrid inflation models have been proposed, depending on the
symmetry of the K\"ahler potential; $R^2$ Starobinsky
model~\cite{Starobinsky:1980te,Mukhanov:1981xt} appears from the
superconformal symmetry~\cite{Buchmuller:2013zfa}, the chaotic regime
emerges from the shift
symmetry~\cite{Buchmuller:2014rfa,Buchmuller:2014dda}, and
$\alpha$-attractor~\cite{Kallosh:2013yoa} comes from their
combination, which is called superconformal
  subcritical hybrid inflation~\cite{Ishiwata:2018dxg}. A generalized
version of the inflation model proposed in
Ref.~\cite{Ishiwata:2018dxg} is intensitively analyzed in
Ref.~\cite{Gunji:2021zit}.  In our paper we study the neutrino mixing
and the dynamics of inflation based on the model in
Ref.~\cite{Gunji:2021zit}. In the framework, an unconventional
neutrino mixing pattern and the CP phases are obtained. We find that
only the inverted hierarchy is consistent with the data and in the
valid parameters the upper limit for the scale of the Majorana mass is
obtained for successful inflation.

This paper is organized as follows. In Sec.~\ref{sec:model}, we
introduce the supergravity model to study. Then neutrino masses and
mixing pattern are discussed in Sec.~\ref{sec:neutrino}. Here we show
the predicted CP phases and the effective neutrino mass for the
neutrinoless double beta decay. In Sec.~\ref{sec:Inflation}, the
dynamics of inflation is studied both analytically and numerically,
especially focusing on the impact of the Majorana mass
terms. Sec.~\ref{sec:conclusion} contains our conclusion.

\section{The model}
\label{sec:model}

\subsection{Brief review of modular symmetry}
The modular symmetry is the geometric symmetry of the two dimensional
torus.  The two dimensional torus $T^2$ is defined by
$\mathbb{C}/\Lambda$, i.e., the complex plane $\mathbb{C}$ divided by
the two dimensional lattice $\Lambda=\{\sum_{i=1}^2n_i\omega_i\, |\,
n_i \in \mathbb{Z}\}$ with basis vectors $\omega_i\in \mathbb{C}$.
Here, the basis vectors are given by $\omega_1=2\pi R$ and $\omega_2=2\pi
R\tau$ with $R\in\mathbb{R}$, and $\tau=\omega_2/\omega_1$ is the
modulus defined in the upper half plane $\mathcal{H}=\{\tau\in
\mathbb{C}\,|\,\mathrm{Im}\tau>0\}$.  The same lattice is constructed
using the basis vectors transformed as
\begin{align}
  \begin{pmatrix}
    \omega_2^\prime\\
    \omega_1^\prime
  \end{pmatrix}
  &=
  \gamma
  \begin{pmatrix}
    \omega_2\\
    \omega_1
  \end{pmatrix}\,,
  \qquad
  \gamma\in SL(2,\mathbb{Z})\,,
\end{align}
where
\begin{align}
  \Gamma
  \equiv
  SL(2,\mathbb{Z})
  \equiv
  \biggl\{
    \gamma=
    \begin{pmatrix}
      a&b\\
      c&d
    \end{pmatrix}
    \,
    \biggl|
    \biggr.
    \,\,
    a,b,c,d\in\mathbb{Z},\ 
    ad-bc=1
  \biggr\}\,.
\end{align}
Then, the modulus is transformed as
\begin{align}
  \tau\to\gamma\tau=\frac{a\tau+b}{c\tau+d}\,.
  \label{eq:Modular_transformation}
\end{align}
Eq.~\eqref{eq:Modular_transformation} is called modular
transformation.  It is seen that the transformation law of $\tau$ is
the same for $\gamma$ and $-\gamma$.  Then, the group of modular
transformation is isomorphic to $PSL(2,\mathbb{Z})\equiv
SL(2,\mathbb{Z})/\pm\mathbb{I}\equiv \bar{\Gamma}$, which is called
modular group.  The modular group is generated by two generators,
\begin{align}
  S
  &=
  \begin{pmatrix}
    0&1\\
    -1&0
  \end{pmatrix},\quad
  T
  =
  \begin{pmatrix}
    1&1\\
    0&1
  \end{pmatrix}\,,
\end{align}
satisfying
\begin{align}
  (ST)^3=S^2=\mathbb{I}\,.
\end{align}
The modulus $\tau$ is transformed by $S$ and $T$ as
\begin{align}
  \begin{split}
    S&:\, \tau\to-1/\tau\,,\\
    T&:\, \tau\to\tau+1\,.
  \end{split}
\end{align}
For a positive integer $N$, the principal congruence subgroup of level $N$ is defined by
\begin{align}
  \Gamma(N)
  &=
  \biggl\{
    \begin{pmatrix}
      a&b\\
      c&d
    \end{pmatrix}
    \in \Gamma \ 
    \biggl|
    \biggr.\,
    \begin{pmatrix}
      a&b\\
      c&d
    \end{pmatrix}
    \equiv
    \begin{pmatrix}
      1&0\\
      0&1
    \end{pmatrix}
    \,(\mathrm{mod}\,N)
  \biggr\}\,.
\end{align}
Note that $\Gamma(1)=\Gamma$ and $-\mathbb{I}\not\in\Gamma(N)$ for
$N>2$.  Then, we introduce the groups $\bar{\Gamma}(N)$ as
$\bar{\Gamma}(2)=\Gamma(2)/\pm\mathbb{I}$ and
$\bar{\Gamma}(N)=\Gamma(N)\, (N>2)$.  The quotient groups
$\Gamma_N\equiv \bar{\Gamma}/\bar{\Gamma}(N)$ are called finite
modular groups and the generators of $\Gamma_N$ have the additional
relation,
\begin{align}
  T^N=\mathbb{I}\,.
\end{align}
It is known that the finite modular groups for $N=2,3,4,5$ are
isomorphic to the non-Abelian discrete groups, $\Gamma_2\simeq S_3$,
$\Gamma_3\simeq A_4$, $\Gamma_4\simeq S_4$, $\Gamma_5\simeq A_5$,
respectively\cite{deAdelhartToorop:2011re}. Under the modular
transformation, the holomorphic functions of $\tau$, called the
modular forms are transformed as
\begin{align}
  f(\tau)
  &\to
  f(\gamma\tau)
  =
  (c\tau+d)^k  \rho(\gamma) f(\tau)\,.
\end{align}
Here $\rho(\gamma)$ is a unitary transformation of $\Gamma_N$ $(N>2)$
and $k$ is non-negative even integer, called modular weight.

In this study, we focus on $\Gamma_3$, i.e., $A_4$.  Then, the modular
forms $Y=(Y_1,Y_2,Y_3)^T$ with representation of $A_4$ triplet and
modular weight $2$ are given by~\cite{Feruglio:2017spp}
\begin{align}
  Y_1
  &=
  \frac{i}{2\pi}
  \biggl[
    \frac{\eta^\prime(\tau/3)}{\eta(\tau/3)} 
    + \frac{\eta^\prime((\tau+1)/3)}{\eta((\tau+1)/3)} 
    + \frac{\eta^\prime((\tau+2)/3)}{\eta((\tau+2)/3)} 
    - \frac{27\eta^\prime(3\tau)}{\eta(3\tau)}
  \biggr]\nonumber
  \,,\\
  Y_2
  &=
  \frac{-i}{\pi}
  \biggl[
    \frac{\eta^\prime(\tau/3)}{\eta(\tau/3)} 
    + \omega^2 \frac{\eta^\prime((\tau+1)/3)}{\eta((\tau+1)/3)} 
    + \omega \frac{\eta^\prime((\tau+2)/3)}{\eta((\tau+2)/3)}
  \biggr]
  \,,\\
  Y_3
  &=
  \frac{-i}{\pi}
  \biggl[
    \frac{\eta^\prime(\tau/3)}{\eta(\tau/3)} 
    + \omega \frac{\eta^\prime((\tau+1)/3)}{\eta((\tau+1)/3)} 
    + \omega^2 \frac{\eta^\prime((\tau+2)/3)}{\eta((\tau+2)/3)}
  \biggr]\nonumber
  \,,
\end{align}
where $\omega=e^{2\pi i/3}$ and $\eta(\tau)$ is the Dedekind eta-function defined by
\begin{align}
  \eta(\tau)
  &=
  q^{1/24}\prod_{i=1}^\infty(1-q^n),\quad q=e^{2\pi i\tau}\,.
\end{align}
The modular forms $Y_i\,(i=1 \text{--} 3)$ satisfy the relation,
\begin{align}
  Y_2^2 + 2 Y_1Y_3 = 0 \,.
  \label{eq:relation_of_Y}
\end{align}
They are transformed under $T$ and $S$ transformations as
\begin{align}
  \begin{split}
    S&:\, Y(-1/\tau)=\tau^2\rho(S)Y\,,\\
    T&:\, Y(\tau+1)=\rho(T)Y\,,
  \end{split}
\end{align}
where
\begin{align}
  \rho(S)
  &=
  \frac{1}{3}
  \begin{pmatrix}
    -1&2&2\\
    2&-1&2\\
    2&2&-1
  \end{pmatrix}\,,
  \quad
  \rho(T)
  =
  \begin{pmatrix}
    1 & &\\
      & \omega & \\
    &&\omega^2
  \end{pmatrix}\,.
\end{align}

It is assumed that the superfields, denoted as $Z^I$, are also
transformed under $A_4$ as
\begin{align}
  Z^I\to(c\tau+d)^{-k_I}\rho^{(I)}(\gamma)Z^I\,,
\end{align}
where the modular weight $k_I$ will be determined later.

\subsection{The superpotential and K\"ahler potential}
\label{sec:WandK}

\begin{table}[tb]
  \centering
  \begin{tabular}{ccccccc}
    \hline
    \parbox[c][7mm][c]{0mm}{}  & $L$ & $E^c=(e^c,\,\mu^c,\,\tau^c)$ & $N^c$ & $H_u$ & $H_d$ & $S_\pm$\\
    \hline
    \parbox[c][7mm][c]{0mm}{}
    U(1) & $0$ & $0$ & $0$ & $0$ & $0$ & $\pm q$\\
    \parbox[c][7mm][c]{0mm}{}
    $A_4$ & $\boldsymbol{3}$ & $(\boldsymbol{1},\,\boldsymbol{1}'',\,\boldsymbol{1}')$ & $\boldsymbol{3}$ & $\boldsymbol{1}$ & $\boldsymbol{1}$ & $\boldsymbol{1}$\\
    \parbox[c][7mm][c]{0mm}{}
    weight & $-k_L$ & $(-k_{e^c},-k_{\mu^c},-k_{\tau^c})$ & $-k_{N^c}$ & $-k_{H_u}$ & $-k_{H_d}$ & $-k_{S_\pm}$\\
    \hline
  \end{tabular}
  \caption{Field contents and their representations and
    modular weights.}
  \label{table:reps_and_weights}
\end{table}

We consider a supergravity model of inflation inspired by modular
$A_4$ symmetry. In addition to the fields in the minimal
supersymmetric standard model (MSSM), we introduce three right-handed
neutrinos $N^c_i\,\,(i=1\text{--}3)$ and two new fields $S_\pm$. Here
$S_\pm$ are gauge singlets under the MSSM gauge but are charged under a
local U(1) with charge $\pm q\,(q>0)$. They play important roles during
inflation.

The superpotential allowed under $A_4$ symmetry is given by 
\begin{align}
  W&\supset W_E+W_N+W_{\rm hyb}\,,
\end{align}
where 
\begin{align}
  W_{E}
  &=\alpha_1 e^cH_d(LY)_{\boldsymbol{1}}
    +\alpha_2 \mu^cH_d(LY)_{\boldsymbol{1}'}
    +\alpha_3 \tau^cH_d(LY)_{\boldsymbol{1}''}\,,
    \label{eq:W_E}\\
  W_{N}
  &=g_1(N^cH_u(LY)_{\boldsymbol{3}\mathrm{s}})_{\boldsymbol{1}}
  +g_2(N^cH_u(LY)_{\boldsymbol{3}\mathrm{a}})_{\boldsymbol{1}}
  \,,
  \label{eq:W_N}\\
  W_{\rm hyb}
  &=\lambda S_+S_-(N^cY)_{\boldsymbol{1}}+\Lambda(N^cN^cY)_{\boldsymbol{1}}\,.
  \label{eq:W_lambda}
\end{align}
The superpotential is similar to one considered in
Ref.~\cite{Gunji:2019wtk}, but it is extended to accommodate the
modular $A_4$ symmetry.  The contents of the fields are listed in
Table~\ref{table:reps_and_weights}. The representation and the modular
weights of the fields are also shown in the table, which is based on
the model considered in
Ref.~\cite{Feruglio:2017spp,Kobayashi:2018scp}; the right-handed
neutrinos and the left-handed lepton doublets
$L_i=(\nu_{Li},l_{Li})^T\,\,(i=e,\mu,\tau)$ are the $A_4$ triplets and
the others are the singlets. Here the three different
  singlet representations $(\mathbf{1},\mathbf{1''},\mathbf{1'})$ are
  assigned to the right-handed charged leptons
  $E^c=(e^c,\mu^c,\tau^c)$, respectively. We will discuss the
assignment of the modular weights soon later.  In the superpotential,
$\alpha_i\,\,(i=1\text{--}3)$, $g_i\,\,(i=1,2)$, and $\lambda$ are
Yukawa coupling constants, and $\Lambda$ determines the mass scale of
the right-handed neutrinos.  By redefinition of the fields, the
parameters $\alpha_i$, $\lambda$, $\Lambda$, and $g_1$ can be taken to
be real without the loss of generality.

For K\"ahler potential, we adopt a class of the canonical
superconformal supergravity model~\cite{Ferrara:2010in}. This type of
model can be extended to the model with a parameter $\alpha$, which
corresponds to the parameter of the superconformal
$\alpha$ attractor model~\cite{Kallosh:2013yoa}. Recently the dynamics
of inflation has been analyzed in a generalized version of the
model~\cite{Gunji:2021zit}. In order to consider the similar inflation
model, we consider the K\"ahler potential based on
Ref.~\cite{Gunji:2021zit}:\footnote{We adopt the Planck unit, where
$M_{\mathrm{pl}}=1$ for the Planck mass $M_{\mathrm{pl}}$.}
\begin{align}
  K &= -3 \log \Bigl( -\frac{\mathcal{N}}{3} \Bigr)\,,
\end{align}
where
\begin{align}
  \mathcal{N}
  &=
  -|Z^0|^n\Bigl(-\frac{\Phi}{3}\Bigr)^\alpha\,,\quad
  -\frac{\Phi}{3}
  =
    1-\sum_I\frac{|Z^I|^2}{|Z^0|^{k_I}}\,.
    \end{align}
Here we have introduced a positive parameter $n$ in order to discuss
the modular weights of the field generically.  We take $Z^0$ as
$Z^0=\bar{Z^0}= 2 \,\mathrm{Im}\,\tau$, In general, $\Phi$ includes
additional terms, such as $(YN^cN^c)_{\mathbf{1}}$ or
$(\bar{Y}\bar{L}YL)_{\mathbf{1}}$~\cite{Feruglio:2017spp,
  Chen:2019ewa}. In the current study we ignore them to focus on the
simple setting.\footnote{This is also motivated by the results given
by Ref.~\cite{Gunji:2021zit}. In the literature it is
  shown such a simple K\"aher potential gives a consistent result with
  the CMB observations.}

The K\"ahler potential is further divided into the modular and matter
parts as $K = K^\tau + K^{\mathrm{m}}$, where
\begin{align}
  K^\tau
  &=
  -3n\log(2\,\mathrm{Im}\,\tau)\,,\quad
  K^{\mathrm{m}}
  =
  -3\alpha\log
  \Bigl(-\frac{\Phi}{3}\Bigr)
    \,.
\end{align}
The matter part is constructed to be modular invariant. Then, under
the modular transformation~\eqref{eq:Modular_transformation}, the
K\"ahler potential is transformed as
\begin{align}
  K
  &\to
  K+3n\bigl[\log(c\tau+d)+\log(c\bar{\tau}+d)\bigr]\,.
\end{align}
In the supergravity the combination of the K\"ahler potential and
superpotential, which is defined by
\begin{align}
  \mathcal{G}
  &=
  K+\log{W}+\log{\overline{W}}\,,
\end{align}
should be modular invariant. Due to the invariance, the transformation of
the superpotential is determined as,
\begin{align}
  \begin{split}
   W&\to e^{i\alpha(\gamma)}(c\tau+d)^{-3n}W\,,\\
  \overline{W}&\to e^{-i\alpha(\gamma)}(c\bar{\tau}+d)^{-3n}\overline{W}\,.
  \end{split}
\end{align}
Namely, its modular weight is $-3n$.  Consequently, the weight of
fields should satisfy
\begin{align}
  \begin{split}
    3n&=
    k_{H_u}+k_{H_d}
    =
    k_{S_+}+k_{S_-}+k_{N^c}-2
    =
    2k_{N^c}-2\\
    &=
    k_{N^c}+k_L+k_{H_u}-2
    =
    k_{E^c}+k_L+k_{H_d}-2
    \,,
  \end{split}
  \label{eq:weights_W}
\end{align}
where $k_{e^c}=k_{\mu^c}=k_{\tau^c}=k_{E^c}$. Those are the
generic conditions for the modular weights in the present model. 
For instance, if we further impose the following conditions:
\begin{align}
  k_{H_u}=k_{H_d}\equiv k_H\,,\quad 
  k_{S_+}=k_{S_-}\equiv k_S\,,
\end{align}
the modular weights are uniquely determined for a given $n$ as
\begin{align}
  k_L=1\,,
  \quad
  k_{N^c}=k_{E^c}=
  2k_S=k_H+1=(3n+2)/2\,.
  \label{eq:weights_eg}
\end{align}
The value of $n$ is a free parameter and it may restrict possible terms
in the K\"ahler potential. However, we do not consider this direction
seriously in the current study.

After the modulus parameter $\tau$ is fixed, we redefine the
chiral superfields as
\begin{align}
  Z^I \to \check{Z}^{I}=Z^I\sqrt{3/|Z^0|^{k_I}}\,.
\end{align}
Accordingly it is convenient to reparametrize the Yukawa couplings
($\alpha_i$, $g_i$, $\lambda$) and $\Lambda$ as
($\check{\alpha}_i$, $\check{g}_i$, $\check{\lambda}$) and
$\check{\Lambda}$ to give rise to the same form of $W_E+W_N+W_{\rm
  hyb}$. Hereafter, we will write the model in this
field basis and omit the `check' symbol for a simple notation unless
otherwise noticed. Namely, we use the superpotential given in
Eqs.~\eqref{eq:W_E}\,--\,\eqref{eq:W_lambda} and the K\"ahler
potential $K^{\rm m}$ of the matter part with the function $\Phi$
\begin{align}
 -\frac{\Phi}{3}=1-\frac{1}{3}\sum_I |Z^{I}|^2\,.
\end{align}
For later calculation, we rewrite $W_{\rm hyb}$ as
\begin{align}
  W_{\rm hyb} = \lambda_i S_+S_-N^c_i + \frac{1}{2}M_{ij} N^c_iN^c_j\,,
  \label{eq:W_hyb}
\end{align}
where
$(\lambda_1,\lambda_2,\lambda_3)=\lambda(Y_1,Y_3,Y_2)$
and
\begin{align}
  M
  &=
  \Lambda
  \begin{pmatrix}
    2Y_1 & -Y_3 & -Y_2\\
    -Y_3 & 2Y_2 & -Y_1\\
    -Y_2 & -Y_1 & 2Y_3
  \end{pmatrix}\,.
  \label{eq:M}
\end{align}

\section{Neutrino masses and mixing pattern}
\label{sec:neutrino}

In this section, we discuss the lepton sector, especially focusing on the
neutrino mixing pattern. From the superpotential, the light neutrinos
acquire masses in the seesaw
mechanism~\cite{Minkowski:1977sc,Yanagida:1979as,Yanagida:1980xy,Gell-Mann:1979vob,Ramond:1979py,Glashow:1979nm}. In
our model, however, $S_+$ obtains a vacuum expectation value (VEV),
denoted as $\langle S_+\rangle$, at the global minimum, which leads to
an unconventional mass matrix for the light
neutrinos~\cite{Gunji:2019wtk}. In addition, the components of the
mass matrices are given in a limited number of parameters due to $A_4$
symmetry, which results in the characteristic pattern of the neutrino
mixing and CP phases in the lepton sector.

\subsection{Neutrino masses and PMNS matrices}

To give the mass matrices of the leptons, it is
  convenient to adopt canonically normalized field basis. The
canonically normalized field $\hat{Z}^{I}$ is obtained by
\begin{align}
  \hat{Z}^I =\sqrt{\alpha}Z^I\,.
\end{align}
Here we have used $-\Phi/3\simeq 1 $ at the global minimum. The
validity of the approximation is guaranteed by $\expval{S_+}\ll 1$,
which will be shown in Sec.~\ref{sec:Inflation}. Accordingly we
absorb the factor $\sqrt{\alpha}$ by introducing
\begin{align}
  (\hat{\alpha}_i,\hat{g}_i,\hat{\lambda}) = \alpha^{-3/2}(\alpha_i,g_i,\lambda)\,,
  ~~~\hat{\Lambda} = \alpha^{-1}\Lambda\,,
\end{align}
to give the same form of the superpotential.

Let us see the neutrino mass matrix. We rewrite Eq.~\eqref{eq:W_lambda} as
\begin{align}
  W_{\rm hyb} = \hat{\lambda}_i \hat{S}_+\hat{S}_-\hat{N}^c_i
  + \frac{1}{2}\hat{M}_{ij} \hat{N}^c_i\hat{N}^c_j\,.
\end{align}
where
$(\hat{\lambda}_1,\hat{\lambda}_2,\hat{\lambda}_3)=\hat{\lambda}(Y_1,Y_3,Y_2)$
and
\begin{align}
  \hat{M}
  &=
  \hat{\Lambda}
  \begin{pmatrix}
    2Y_1 & -Y_3 & -Y_2\\
    -Y_3 & 2Y_2 & -Y_1\\
    -Y_2 & -Y_1 & 2Y_3
  \end{pmatrix}\,.
\end{align}
Then, the mass matrix of the neutrinos in the basis of $(N^c_1, N^c_2,
N^c_3, \tilde{S}_-, \nu_{Le}, \nu_{L\mu}, \nu_{L\tau})$\footnote{We
use the same notation for the fermionic part as the chiral superfield
for $N^c_i$ and the leptons in the MSSM while $\tilde{S}_-$ is the
fermionic part of $S_-$.}  is given by
\begin{align}
  \mathcal{M}
  &=
  \begin{pmatrix}
    \tilde{M} &   \tilde{M}_D\\
    \tilde{M}_D^T & O
  \end{pmatrix}
  \,.
\end{align}
Here $\tilde{M}$ and $\tilde{M}_D$ are $4\times4$ and $4\times3$
matrices, respectively, written by
\begin{align}
    \tilde{M}
  &=
  \hat{\lambda} \langle \hat{S}_+\rangle
  \left(
  \begin{array}{@{}c:c@{}}
    \hat{M}/\hat{\lambda} \langle \hat{S}_+\rangle &
    \begin{matrix} 
    Y_1 \\
    Y_3 \\
    Y_2 
  \end{matrix}\\
    \hdashline
    \begin{matrix}
      Y_1 &
      Y_3 &
      Y_2 
    \end{matrix}
  & 0
  \end{array}
  \right)\,,
  \label{eq:M_til}\\
  \tilde{M}_D
  &=
  \langle \hat{H}_u^0\rangle
  \begin{pmatrix}
    2\hat{g}_1Y_1&(-\hat{g}_1+\hat{g}_2)Y_3&(-\hat{g}_1-\hat{g}_2)Y_2\\
    (-\hat{g}_1-\hat{g}_2)Y_3&2\hat{g}_1Y_2&(-\hat{g}_1+\hat{g}_2)Y_1\\
    (-\hat{g}_1+\hat{g}_2)Y_2&(-\hat{g}_1-\hat{g}_2)Y_1&2\hat{g}_1Y_3\\
    0&0&0
  \end{pmatrix}\,,
  \label{eq:M_til_D}
  \end{align}
where $\langle \hat{H}_u^0\rangle$ is the VEV of up-type neutral
Higgs. We note that $\hat{\lambda}\expval*{\hat{S}_+}$ corresponds to
the scale of the inflaton mass, which will be shown
later.
Consequently, the light neutrinos mass matrix is obtained by the seesaw
mechanism as
\begin{align}
  M_\nu
  &=
  -\tilde{M}_D^T\tilde{M}^{-1}\tilde{M}_D\,.
  \label{eq:M_nu}
\end{align}
Using Eq.~\eqref{eq:relation_of_Y}, it is straightforward to obtain
the mass matrix $M_\nu$ as
\begin{align}
  \frac{M_{\nu11}}{m_{\nu 0}}
  &=
  12\hat{Y}_2^2
  \bigl[
    -2(3+5\hat{g})+(-3+\hat{g})\hat{Y}_2^3
  \bigr]
  \,,\nonumber\\
  \frac{M_{\nu22}}{m_{\nu 0}}
  &=
  \frac{
    4(2+\hat{Y}_2^3)
    \bigl[
      (1+\hat{g})\hat{Y}_2^6+16\hat{Y}_2^3+8(-1+\hat{g})
    \bigr]
  }
  {\hat{Y}_2}\,,\nonumber\\
  \frac{M_{\nu12}}{m_{\nu 0}}
  &=
  \frac{M_{\nu21}}{m_{\nu 0}}
  =
  -\frac{M_{\nu33}}{2m_{\nu 0}}
  =
  (-4+\hat{Y}_2^3)
  \bigl[ 
    4(3+\hat{g})+(-3+5\hat{g})\hat{Y}_2^3
  \bigr]
  \,,\\
  \frac{M_{\nu13}}{m_{\nu 0}}
  &=
  \frac{M_{\nu31}}{m_{\nu 0}}
  =
  \hat{Y}_2
  \bigl[
    (-3+\hat{g})(16+\hat{Y}_2^6)
    +4(6-11\hat{g})\hat{Y}_2^3
  \bigr]\,,\nonumber\\
  \frac{M_{\nu23}}{m_{\nu 0}}
  &=
  \frac{M_{\nu32}}{m_{\nu 0}}
  =
  \frac{
    -64(1+\hat{g})
    -48(2+\hat{g})\hat{Y}_2^3
    -24(-1+\hat{g})\hat{Y}_2^6
    +(1+\hat{g})\hat{Y}_2^9
  }
  {2\hat{Y}_2}
  \,,\nonumber
\end{align}
where
\begin{align}
  \begin{split}
   &~~~~~~~~~ m_{\nu 0}
  \equiv
  \frac{(-1+\hat{g})}{(8+\hat{Y}_2^3)^2}
  \frac{\hat{g}_1^2 Y_3 \langle \hat{H}_u\rangle^2}{\hat{\Lambda}}
  \,, \\
  \hat{Y}_1
  &\equiv
  Y_1/Y_3\,,\quad
  \hat{Y}_2
  \equiv
  Y_2/Y_3\,, \quad
  \hat{g}
  \equiv
  \hat{g}_2/\hat{g}_1\,.
  \end{split}
\end{align}

It is worth notifying that the mass matrix is independent of
$\hat{\lambda} \expval*{\hat{S}_+}$. Therefore, the scale of the
inflaton mass is not affected by the observations of the neutrino
sector. Finally $M_\nu$ is diagonalized by a unitary matrix $U_\nu$ as
\begin{align}
  \mathrm{diag}(m_1,m_2,m_3)=U_\nu^TM_\nu U_\nu\,.
\end{align}
The obtained neutrino masses are then compared with the observed
values shown in Table~\ref{table:Nufit5.0}. We notice that the
lightest neutrino mass becomes zero since rank of $M_\nu$ is
two~\cite{Gunji:2019wtk}.
Therefore, by imposing the condition in
  which the neutrino mass squared differences are within $3\sigma$
  range of the observed values, the cosmological
upper bound on the sum of light neutrino
masses~\cite{Aghanim:2018eyx,Vagnozzi:2017ovm}
\begin{align}
  \sum_i m_i\le 120\, \mathrm{meV}\,,
\end{align}
is automatically satisfied in our model for both the normal hierarchy
(NH) and inverted hierarchy (IH).

\begin{table}[tbp]
  \centering
  \begin{tabular}{ccc}
    \hline
    \parbox[c][7mm][c]{0mm}{} & Normal Hierarchy ($3\sigma$ range)&Inverted Hierarchy ($3\sigma$ range)\\
    \hline
    \parbox[c][7mm][c]{0mm}{}$\sin^2\theta_{12}$&$0.269\text{\,--\,}0.343$&$0.269\text{\,--\,}0.343$\\
    \parbox[c][7mm][c]{0mm}{}$\sin^2\theta_{23}$&$0.415\text{\,--\,}0.616$&$0.419\text{\,--\,}0.617$\\
    \parbox[c][7mm][c]{0mm}{}$\sin^2\theta_{13}$&$0.02032\text{\,--\,}0.02410$&$0.02052\text{\,--\,}0.02428$\\
    \parbox[c][7mm][c]{0mm}{}$\displaystyle \Delta m_{21}^2/{10^{-5}\,\mathrm{eV}^2}$&$6.82\text{\,--\,}8.04$&$6.82\text{\,--\,}8.04$\\
    \parbox[c][7mm][c]{0mm}{}$\displaystyle \Delta m_{3l}^2/{10^{-3}\,\mathrm{eV}^2}$&$2.435\text{\,--\,}2.598$&$-2.581\text{\,--\,}-2.414$\\
    \hline
  \end{tabular}
  \caption{Observed data of the neutrino mass squared differences and
    the neutrino mixing angles from NuFIT
    5.0~\cite{Esteban:2020cvm}. Here, $\Delta m_{3l}^2=\Delta
    m_{31}^2>0$ (NH) and $\Delta m_{3l}^2=\Delta m_{32}^2<0$ (IH).}
  \label{table:Nufit5.0}
\end{table}

Another observable is the PMNS matrix, which is defined as $U \equiv
U_l^\dag U_\nu$. Here $U_l$ is a unitary matrix that diagonalize the
charged lepton mass matrix $M_E$. $M_E$ is the same as one studied in
Ref.~\cite{Kobayashi:2018scp}.
The result is parametrized in terms of three mixing angles
$\theta_{12}$, $\theta_{13}$, $\theta_{23}$, a Dirac phase
$\delta_{\mathrm{CP}}$, and a Majorana phase $\alpha_{21}$;
\begin{align}
  U
  &=
   \begin{pmatrix}
     c_{12}c_{13}&s_{12}c_{13}&s_{13}e^{-i\delta_{\mathrm{CP}}}\\
     -s_{12}c_{23}-c_{12}s_{23}s_{13}e^{i\delta_{\mathrm{CP}}}&c_{12}c_{23}-s_{12}s_{23}s_{13}e^{i\delta_{\mathrm{CP}}}&s_{23}c_{13}\\
     s_{12}s_{23}-c_{12}c_{23}s_{13}e^{i\delta_{\mathrm{CP}}}&-c_{12}s_{23}-s_{12}c_{23}s_{13}e^{i\delta_{\mathrm{CP}}}&c_{23}c_{13}
   \end{pmatrix}
   \begin{pmatrix}
     1&&\\
     &e^{i\frac{\alpha_{21}}{2}}&\\
     &&1
   \end{pmatrix}\,,
\end{align}
where $s_{ij}=\sin \theta_{ij}$ and $c_{ij}=\cos \theta_{ij}$. The
mixing angles are determined by the observations and
  they are summarized in Table~\ref{table:Nufit5.0}. We note that
there is only one Majorana phase since one of the light neutrinos is
massless.  Then, the invariant quantities regarding the CP phases are
given by~\cite{Jarlskog:1985ht, Bilenky:2001rz, Nieves:2001fc,
  Aguilar-Saavedra:2000jom, Girardi:2016zwz}
\begin{align}
  J_{CP}
  &=
  \mathrm{Im}\,
  \bigl[
    U_{e1}U_{\mu2}U_{e2}^*U_{\mu1}^*
  \bigr]
  =
  s_{23}c_{23}s_{12}c_{12}s_{13}c_{13}^2\sin\delta_{CP} \,,
  \label{eq:J_CP}\\
  I_1
  &=
  \mathrm{Im}\,
  \bigl[
    U_{e1}^*U_{e2}
  \bigr]
  =
  s_{12}c_{12}c_{13}^2\sin(\alpha_{21}/2)
  \,,\\
  I_2
  &=
  \mathrm{Im}\,
  \bigl[
    U_{e1}^*U_{e3}
  \bigr]
  =
  c_{12}s_{13}c_{13}\sin(-\delta_{CP})\,.
\end{align}
In addition, the following relations are useful to determine the CP
phases~\cite{Okada:2021qdf}:
\begin{align}
  \cos\delta_{CP}
  &=
  \frac{|U_{\tau1}|^2-s_{12}^2s_{23}^2-c_{12}^2c_{23}^2s_{13}^2}
  {-2c_{12}s_{12}c_{23}s_{23}s_{13}}
  \,,\\
  \mathrm{Re}\,\bigl[U_{e1}^*U_{e2}\bigr]
  &=
  c_{12}s_{12}c_{13}^2\cos(\alpha_{21}/2)
  \,,\\
  \mathrm{Re}\,\bigl[U_{e1}^*U_{e3}\bigr]
  &=
  c_{12}s_{13}c_{13}\cos(-\delta_{CP})\label{ReUe1Ue3}
  \,.
\end{align}

\subsection{Observational consequences}

Based on the arguments in the previous subsection, we compute the
mixing angles and CP phases for a given set of parameters.  The
relevant parameters are $\tau$ and $\hat{g}$ and we parametrize
them as
\begin{align}
    \tau=\mathrm{Re}\,\tau+i\,\mathrm{Im}\,\tau\,,\quad
    \hat{g}=ge^{i\phi_g}\,,
\end{align}
where $g=|\hat{g}|$ and $\phi_g$ is argument of $\hat{g}$.
Considering the fundamental domain $\mathcal{F}$ of $\bar{\Gamma}$ for
$\tau$,
\begin{align}
  \mathcal{F}
  &=
  \Bigl\{
    \tau\in \mathcal{H}
    \,\,\Bigl|\Bigr.
    \,\,
    |\mathrm{Re}\,\tau|\le\frac{1}{2},\,\, |\tau|\ge 1
  \Bigr\}	\,,
\end{align}
we scan the following parameter ranges,
\begin{align}
  \begin{split}
    &|\tau|\ge 1,\quad|\mathrm{Re}\,\tau|\le \frac{1}{2},
    \quad\mathrm{Im}\,\tau\le 1.65\,,
    \\
    &0\le g\le 6.5,\quad |\phi_g|\le180^\circ\,.
  \end{split}
\end{align}
We use the 3$\sigma$ data of the mixing angles and mass squared
differences from NuFIT 5.0~\cite{Esteban:2020cvm},
which are listed in Table~\ref{table:Nufit5.0}. In
the analysis, we take the VEVs of the up- and down-type Higgs bosons
as $v/2$ $(v\simeq 246.7\,\mathrm{GeV})$ by considering so-called
high-scale SUSY to give $125\,\mathrm{GeV}$ Higgs
mass~\cite{Giudice:2004tc,Giudice:2011cg}.

\begin{figure}[tbp]
  \begin{minipage}[t]{0.5\hsize}
  \centering
  \includegraphics[keepaspectratio,width=7.8cm]{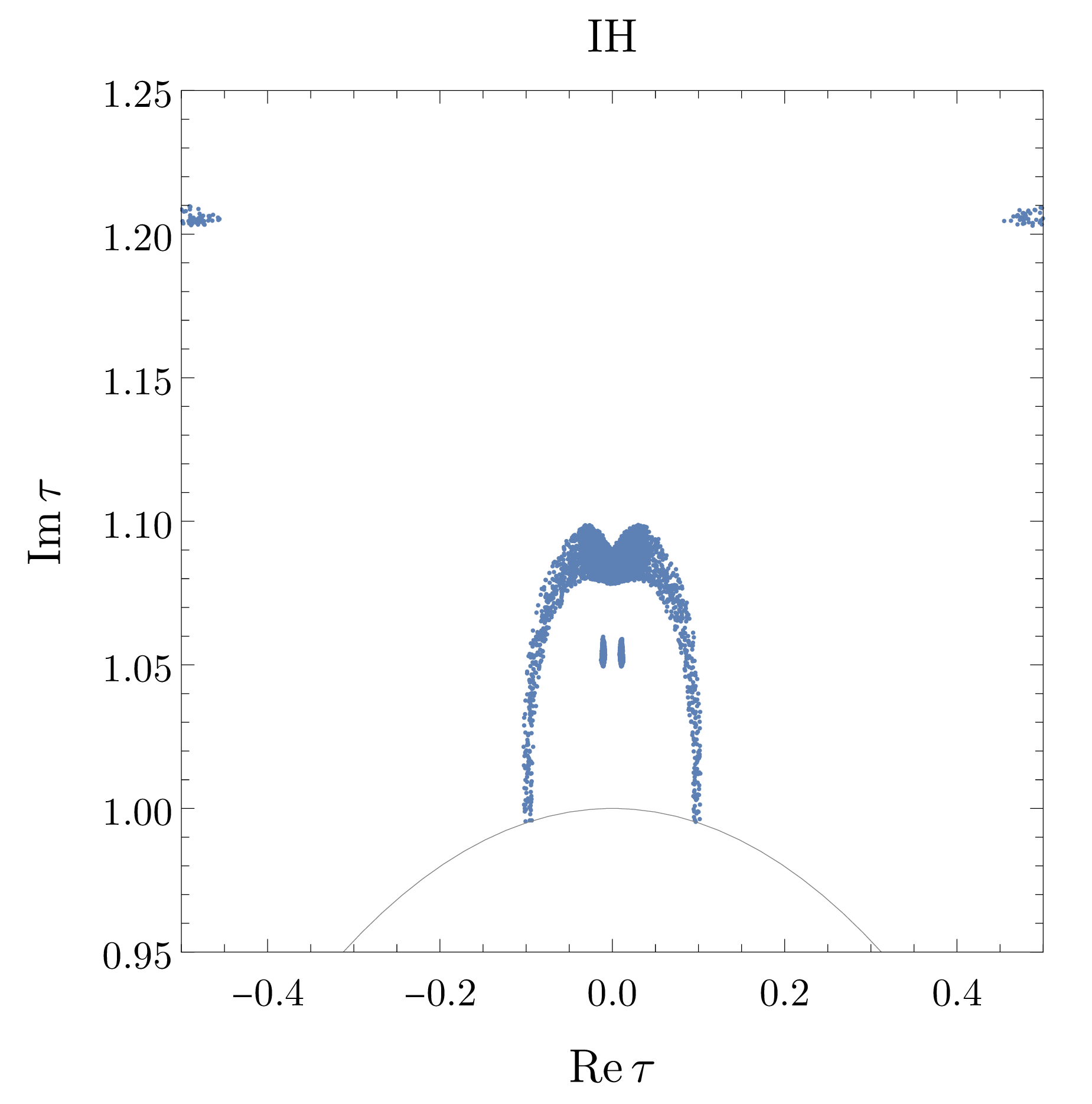}
  \end{minipage}
  \begin{minipage}[t]{0.5\hsize}
  \centering
  \includegraphics[keepaspectratio,width=7.8cm]{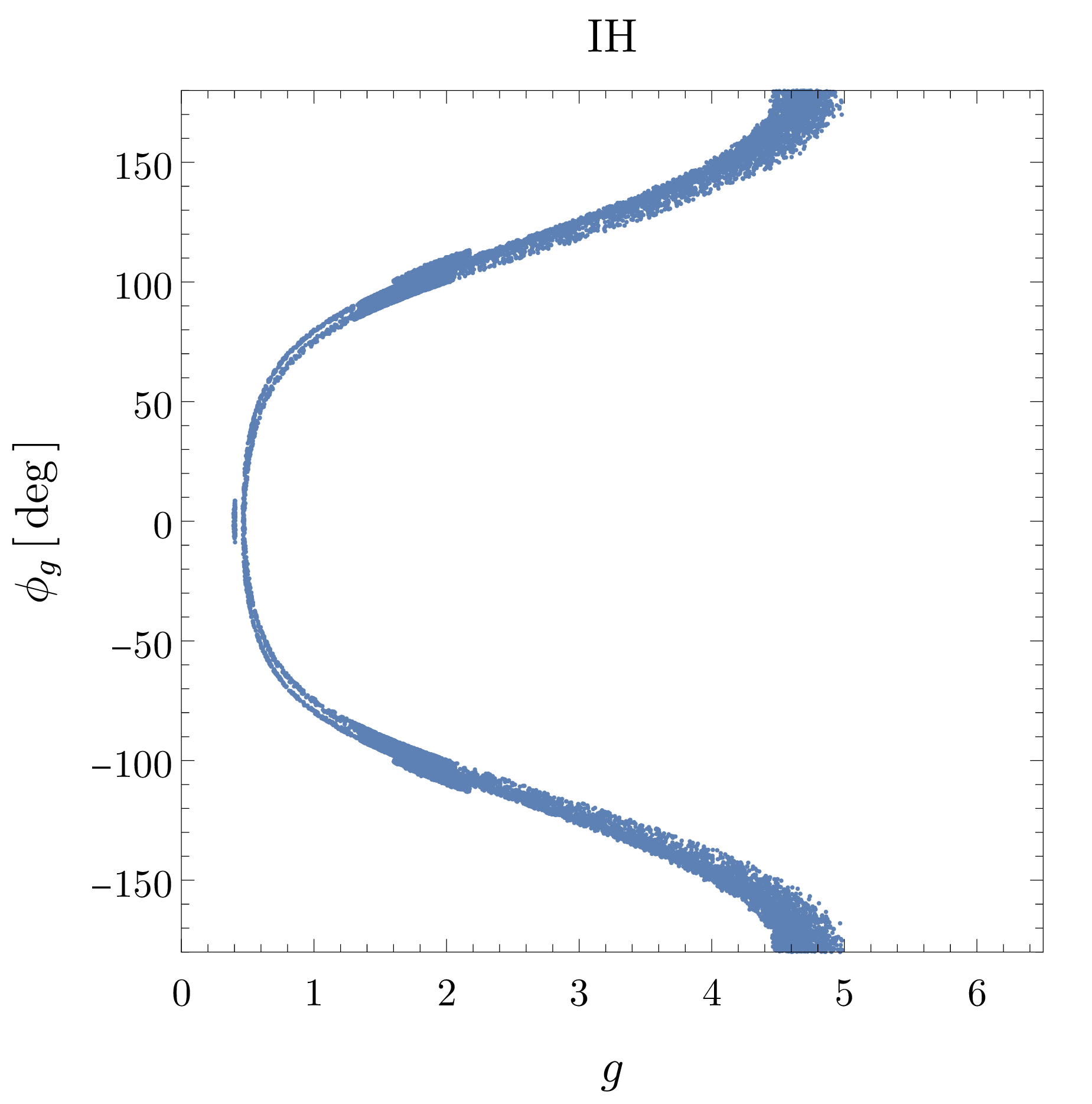}
  \end{minipage}
  \caption{Allowed region of $\tau$ (left) and $\hat{g}=ge^{i\phi_g}$
    (right) for inverted hierarchy. Each dot indicates the value that
    is consistent with the $3\sigma$ data of the neutrino oscillation
    experiments~\cite{Esteban:2020cvm} summarized in
    Table~\ref{table:Nufit5.0}.}
  \label{fig:tau_g}
\end{figure}

First of all, we have found no allowed region for the
  NH case.  Fig.~\ref{fig:tau_g} shows the allowed regions for the
  $\tau$ and $\hat{g}$ for the IH case.
We found a specific pattern of the allowed values of $\tau$ and
$\hat{g}$ is seen.  The allowed value of $\tau$ is limited in
$1.0\lesssim\mathrm{Im}\,\tau\lesssim1.1$ and
$|\mathrm{Re}\,\tau|\lesssim 0.1$, or $\mathrm{Im}\,\tau\sim1.2$ and
$|\mathrm{Re}\,\tau|\sim 0.5$.  This is a different feature compared
to the previous work, e.g., Ref.~\cite{Kobayashi:2018scp}, where the
allowed values of $\tau$ is distributed more widely.
This difference comes from the unconventional pattern
  of the active neutrino mass matrix \eqref{eq:M_nu} given by
  Eqs.\,\eqref{eq:M_til} and \eqref{eq:M_til_D}, which comes from
  additional Yukawa couplings between the right-handed neutrinos and
  $S_+$ in Eq.~\eqref{eq:W_hyb} and the VEV of $S_+$.  Regarding
$\hat{g}$, $g$ is distributed as $0.4\lesssim g\lesssim 5$. The phase
$\phi_g$ can take any value in $[-180^\circ,180^\circ]$, but it is
given by a smooth function of $g$. Such behavior is in contrast to the
results in Ref.~\cite{Kobayashi:2018scp}; in the literature, the
allowed region for both $g$ and $\phi_g$ are more restricted.
\begin{figure}[tbp]
  \begin{minipage}[t]{0.5\hsize}
  \centering
  \includegraphics[keepaspectratio,width=7.8cm]{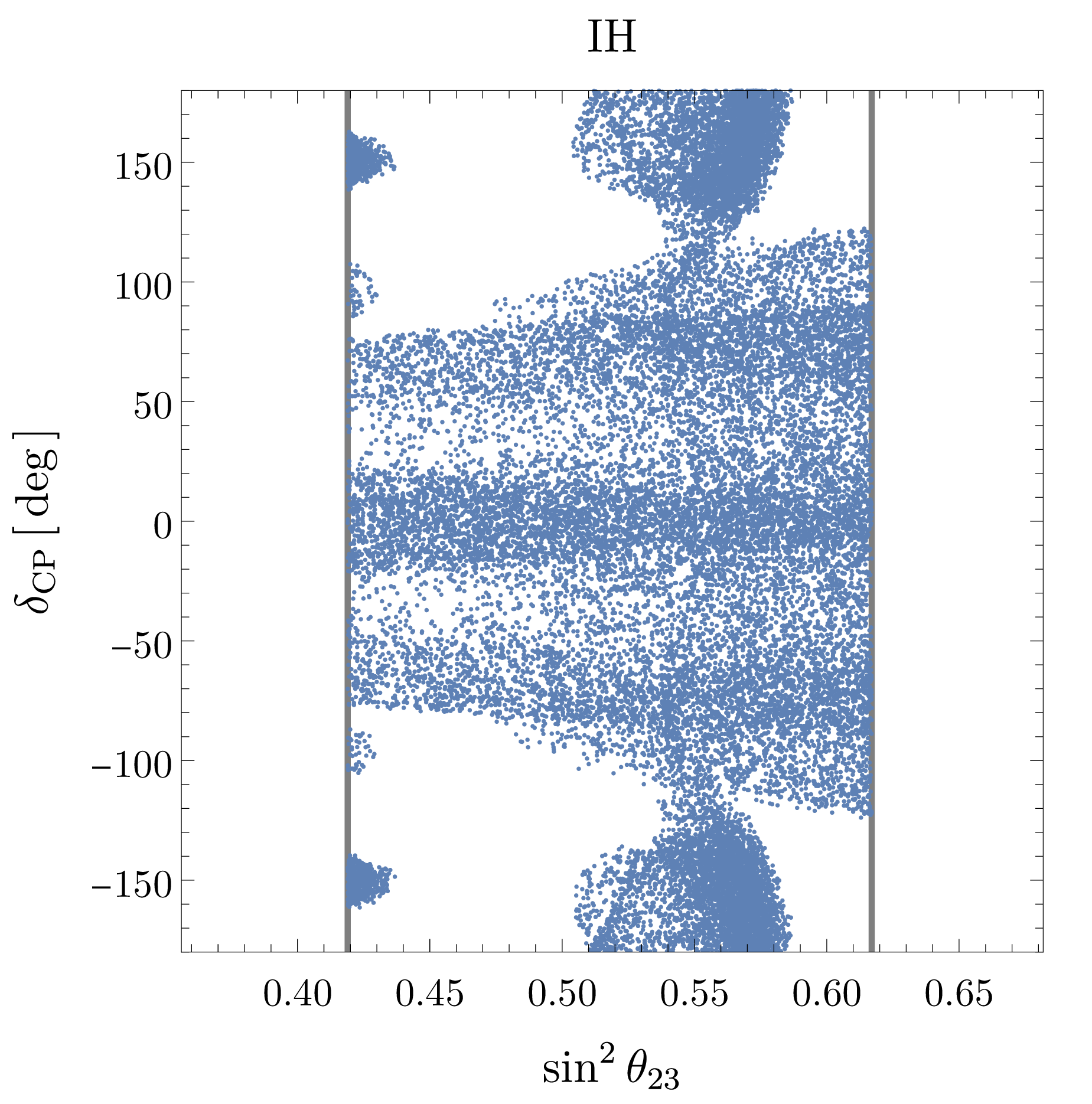}
  \end{minipage}
  \begin{minipage}[t]{0.5\hsize}
  \centering
  \includegraphics[keepaspectratio,width=7.8cm]{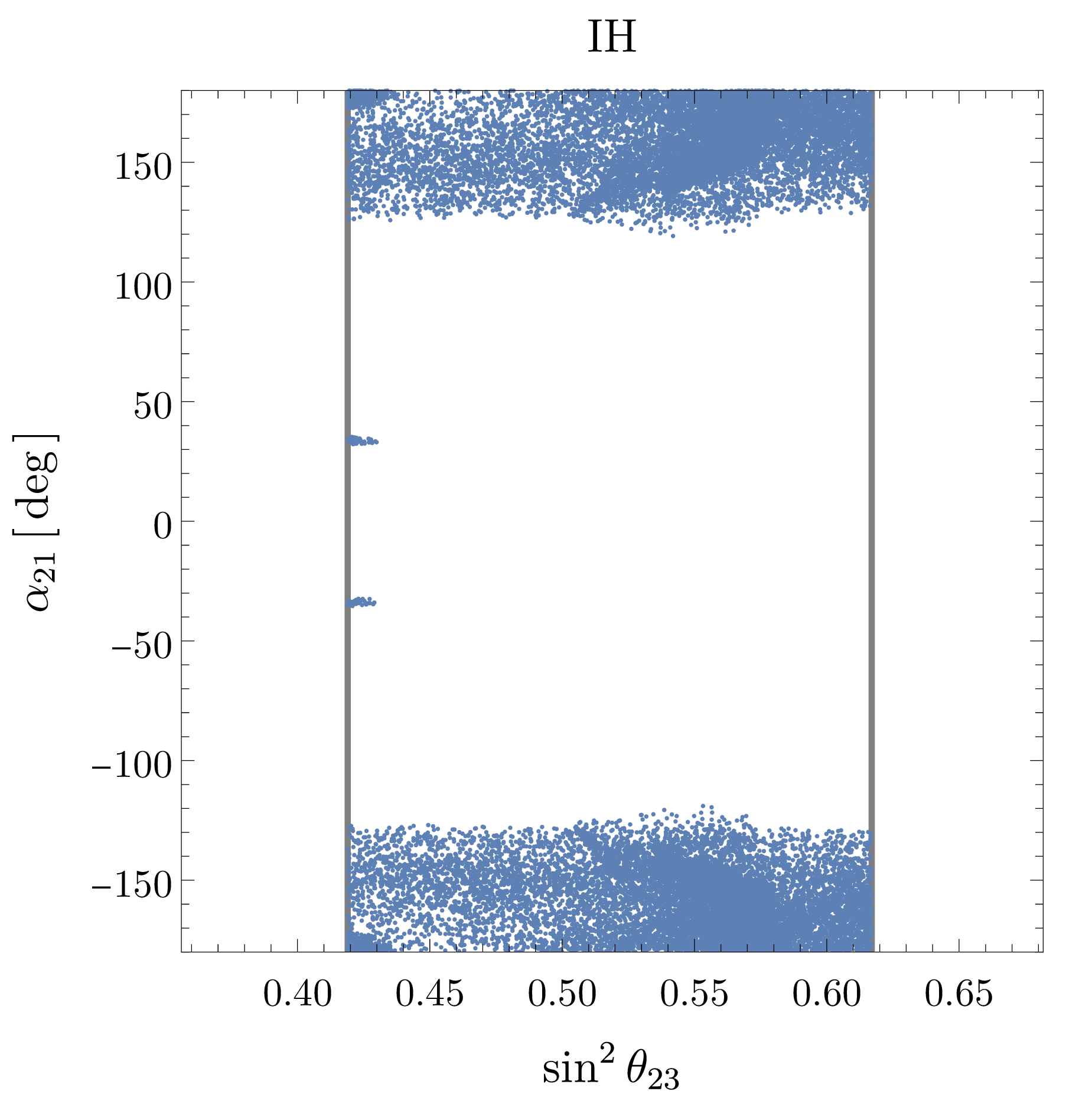}
  \end{minipage}\\
  \begin{center}
  \begin{minipage}[t]{0.5\hsize}
  \centering
  \includegraphics[keepaspectratio,width=7.8cm]{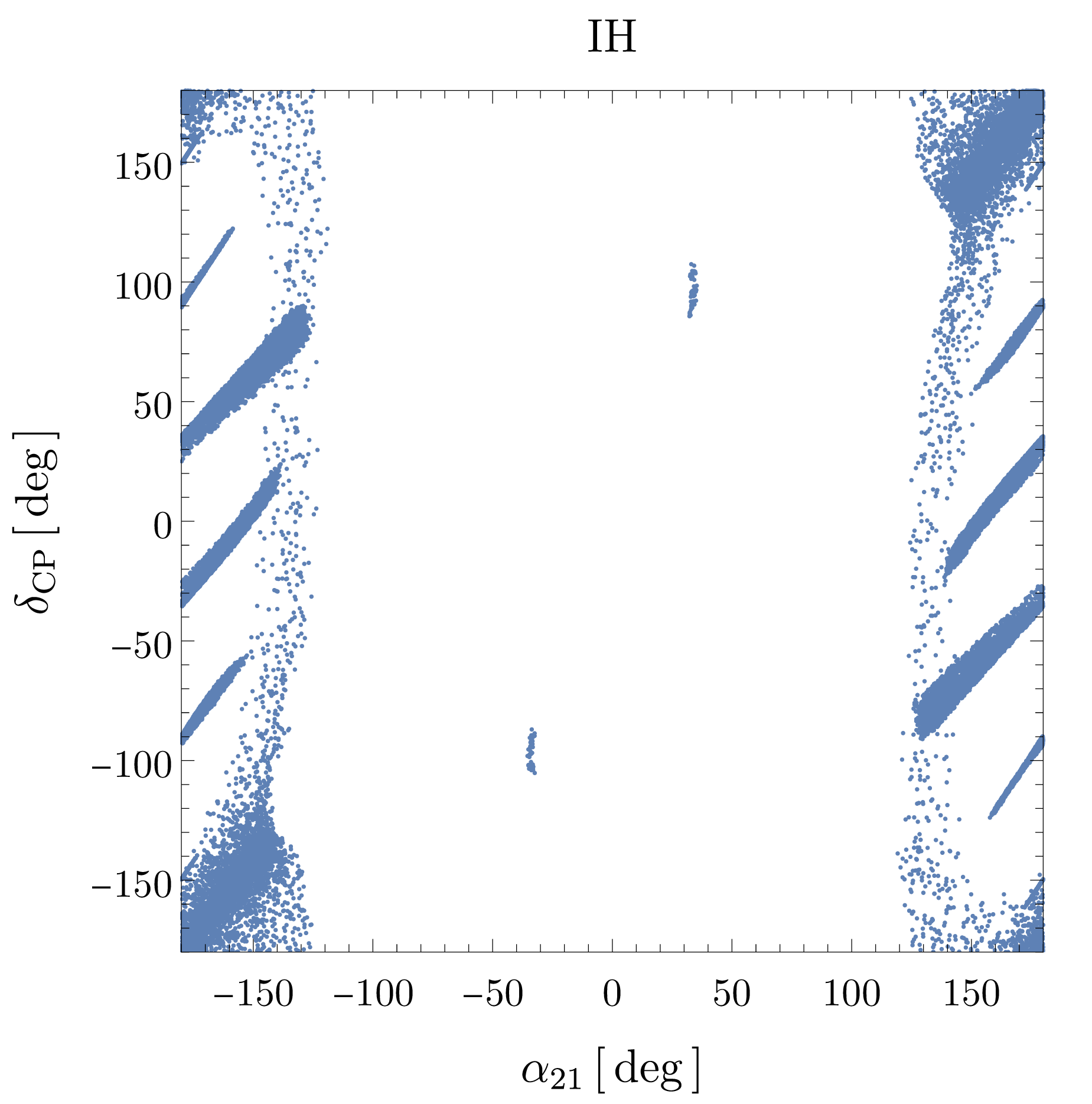}
  \end{minipage}
  \end{center}
  \caption{Predicted CP phases $\delta_{\rm CP}$ and $\alpha_{21}$
    from the allowed parameter sets given in
    Fig.~\ref{fig:tau_g}. The correlations between
    $\sin^2\theta_{23}$ and the CP phases (top-left and top-right) and
    the correlation between CP phases (bottom) are shown.  The gray solid
    lines show the boundaries of the $3\sigma$ region of the
    experimental data.}
  \label{fig:preds}
\end{figure}

The discovered sets of the parameters lead to a new prediction for the CP
phases, which is shown in Fig.~\ref{fig:preds}. In the plot we show
the correlation between $\sin^2\theta_{23}$ and the CP phases. For
$\delta_{\rm CP}$, we found 
$|\delta_{\rm CP}|\lesssim 120^\circ$
for any
value of $\sin^2\theta_{23}$ in the 3$\sigma$ range.  If
$\sin^2\theta_{23}\simeq 0.55$, on the other hand,
then $\delta_{\rm CP}$ can take any values. Regarding $\alpha_{21}$,
we found 
$|\alpha_{21}|\sim 120\text{--}180^\circ$
for any value of $\sin^2\theta_{23}$
in 3$\sigma$ range. An exception is 
$|\alpha_{21}|\sim 35^\circ$
for
$\sin^2\theta_{23}\simeq0.42\text{--}0.43$. Finally, we found no
specific correlation between $\delta_{\rm CP}$ and $\alpha_{21}$,
i.e., 
$|\alpha_{21}|\sim 120\text{--}180^\circ$
is predicted while various value of
$\delta_{\rm CP}$ is possible.  To summarize, only the IH is allowed
and Fig.~\ref{fig:preds} is the prediction for the CP phases in our
model.

\begin{figure}[tbp]
  \begin{minipage}[t]{0.5\hsize}
  \centering
  \includegraphics[keepaspectratio,width=7.8cm]{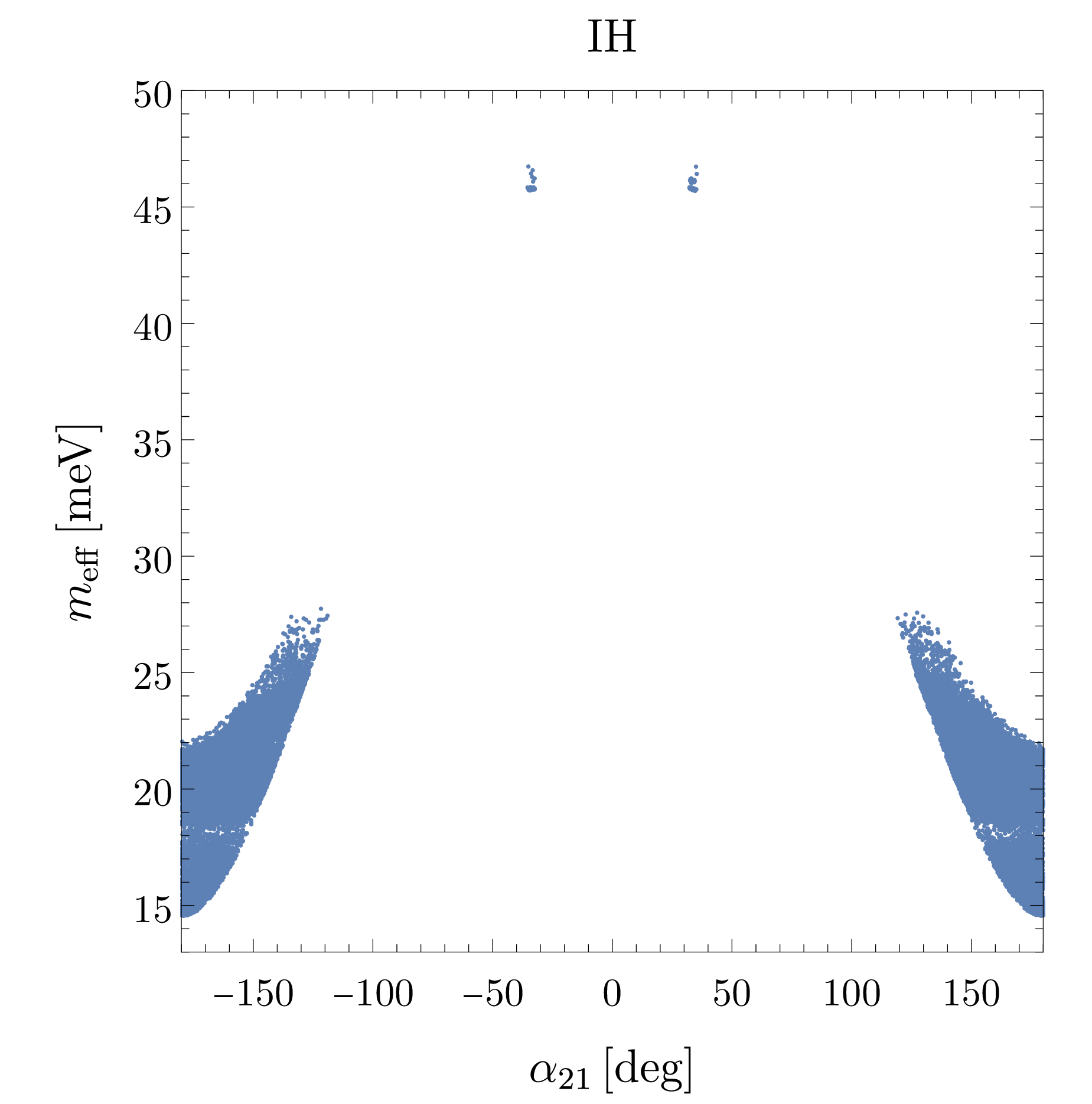}
  \end{minipage}
  \begin{minipage}[t]{0.5\hsize}
  \centering
  \includegraphics[keepaspectratio,width=7.8cm]{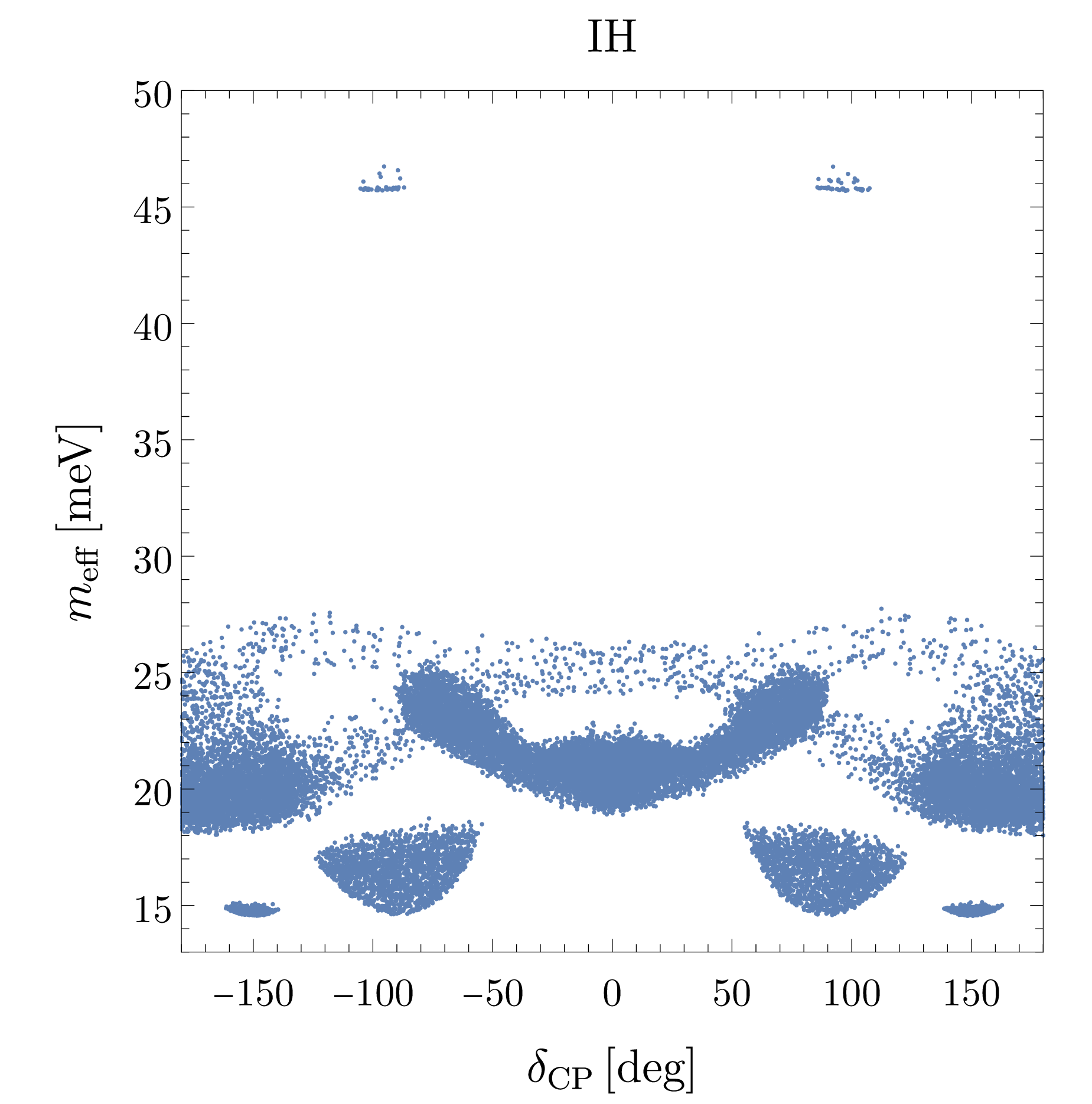}
  \end{minipage}
  \caption{Predicted effective neutrino mass of neutrinoless double
    beta decay as a function of the Majorana CP phase (left) and the
    Dirac CP phase (right), calculated from the results given in
    Fig.~\ref{fig:preds}}
  \label{fig:m_eff}
\end{figure}

Another observable consequence of this model is the neutrinoless
double beta decay $(A, Z) \to (A, Z+2) + 2 e^{-}$, which is a lepton
number violating process in low-energy phenomena. See, e.g.,
Ref.~\cite{Pas:2015eia} for review.  The decay rate of the process is
proportional to the effective neutrino mass-squared, which is given by
\begin{align}
  m_{\rm eff} =\Bigl|\sum_{i}m_i U_{ei}^2 \Bigr|\,.
\end{align}
The current upper limit on the effective mass of
neutrinoless double beta decay from KamLAND-Zen is
36--156~meV~\cite{KamLAND-Zen:2022tow}. Note that the experimental
upper bound on the effective mass is affected by the uncertainty of
the nuclear matrix elements in the decay process.  In the present
model, one of the light neutrinos is massless and we have seen that
the IH is allowed. Therefore, the effective neutrino mass is given
by~\cite{Asaka:2018hyk}
\begin{align}
  m_{\rm eff}^2=c_{13}^4(m_1^2c_{12}^4+m_2^2s_{12}^4+
  2m_1m_2c_{12}^2s_{12}^2\cos\alpha_{21})\,.
  \label{eq:m_eff}
\end{align}
The result is shown in Fig.~\ref{fig:m_eff}. We found $14~{\rm
  meV}\lesssim m_{\rm eff}\lesssim 28~{\rm meV}$ for the most
parameter sets, where $\alpha_{21}\sim \pm 180^\circ$.  One can see
that it becomes as large as around 45--47~meV for $\alpha_{21}\sim \pm
35^\circ$, which can be understood from
Eq.~\eqref{eq:m_eff}. Namely the term proportional to
  $\cos \alpha_{21}$ is constructive in this case. The resultant
value $\order{10}~{\rm meV}$ of $m_{\rm eff}$ is the same as one shown
in Ref.~\cite{Asaka:2018hyk}. Since the model predicts a relatively
large effective mass, neutrinoless double beta decay events may be
observed in future experiments.  If the CP phases are constrained, for
instance, by considering the leptogenesis~\cite{Fukugita:1986hr} in
the present framework, then $m_{\rm eff}$ may be more
restrictive. That would be another future work to pursue (see also the
discussion in Sec.~\ref{sec:inf_num}).

\section{Inflation}
\label{sec:Inflation}

$W_{\rm hyb}$ induces the $D$-term hybrid inflation.
Motivated by the previous
studies~\cite{Ishiwata:2018dxg,Gunji:2021zit}, we consider the
subcritical regime of hybrid inflation, where one of right-handed
sneutrinos plays the role of the inflaton field. In the present case,
however, the additional Majorana mass terms may disturb the dynamics
during inflation. We will derive the inflaton potential and see how
the additional terms affect the dynamics.
In this section, we examine the impact of the heavy Majorana masses on
the subcritical hybrid inflation scenario.  To be concrete,
we assume
\begin{align}
  |M_{ij}|\ll |\lambda_i\langle S_+\rangle|\,.
  \label{eq:assumption}
\end{align}
Under the assumption, we expect that the same inflation model is
obtained as Ref.~\cite{Gunji:2021zit}. We will derive the upper bound
for $M_{ij}$ quantitatively.

\subsection{New field basis}

We start with the basis given in the last paragraph of
Sec.~\ref{sec:WandK}. At the global minimum, the mass matrix squared
$M_{\mathrm{sc}}^2$ of the scalar sector in the basis
$(\tilde{N}^c_1,\tilde{N}^c_2,\tilde{N}^c_3,S_-)$\footnote{$\tilde{N}^c_i$
and $S_\pm$ are scalar components of the chiral superfield $N^c_i$ and
$S_\pm$, respectively.} is obtained as
\begin{align}
  M_{\mathrm{sc}}^2
  &=
  \tilde{M}^\dag\tilde{M}\,,
\end{align}
where $\tilde{M}$ is equal to Eq.~\eqref{eq:M_til} without `hat'
symbol.  Under the condition~\eqref{eq:assumption}, the mass matrix is
approximated as
\begin{align}
  M_{\mathrm{sc}}^2
  &=
  \langle S_+\rangle^2
  \begin{pmatrix}
    |\lambda_1|^2 & \lambda_2\lambda_1^* & \lambda_3 \lambda_1^* & 0\\
    \lambda_1\lambda_2^* & |\lambda_2|^2 & \lambda_3\lambda_2^* & 0\\
    \lambda_1\lambda_3^* & \lambda_2\lambda_3^* & |\lambda_3|^2 & 0\\
    0 & 0 & 0 & \tilde{\lambda}^2
  \end{pmatrix}
  +\mathcal{O}(M_{ij}\lambda_k\langle S_+\rangle)
  \,,
\end{align}
where
\begin{align}
  \tilde{\lambda}\equiv\sqrt{\sum_i |\lambda_i|^2}
\end{align}
and it can be approximately diagonalized by a unitary matrix $V$ as
\begin{align}
  V^\dag M_{\mathrm{sc}}^2 V
  &=
  \langle S_+\rangle^2
  \begin{pmatrix}
    0&&&\\
    &0&&\\
    &&\tilde{\lambda}^2&\\
    &&&\tilde{\lambda}^2
  \end{pmatrix}
  +
  \mathcal{O}(M_{ij}\lambda_k\langle S_+\rangle)\,,
  \end{align}
  \begin{align}
  V
  &\equiv
  \left(
  \begin{array}{@{}c:c@{}}
    W & 
  \begin{matrix} 
    0 \\
    0 \\
    0 
  \end{matrix}\\
    \hdashline
  \begin{matrix}
    0 &
    0 &
    0 
  \end{matrix}
      & \frac{\lambda_3^*}{|\lambda_3|}
  \end{array}
  \right)\,,
  \quad
  W
  \equiv
  \begin{pmatrix}
    \frac{-\lambda_3|\lambda_1|}{\lambda_1 \sqrt{\sum_{j\neq2}|\lambda_j|^2}} &
    \frac{-\lambda_2\lambda_1^*}{\tilde{\lambda}\sqrt{\sum_{j\neq2}|\lambda_j|^2|}} & 
    \frac{|\lambda_3|\lambda_1^*}{\lambda_3^*\tilde{\lambda}} 
    \\
    0 & 
    \sqrt{\frac{\sum_{j\neq2}|\lambda_j|^2}{\tilde{\lambda}^2}} & 
    \frac{|\lambda_3|\lambda_2^*}{\lambda_3^*\tilde{\lambda}}  
    \\
    \frac{|\lambda_1|}{\sqrt{\sum_{j\neq2}|\lambda_j|^2}} & 
    \frac{-\lambda_2\lambda_3^*}{\tilde{\lambda}\sqrt{\sum_{j\neq2}|\lambda_j|^2}} & 
    |\lambda_3|/\tilde{\lambda}
  \end{pmatrix}\,.
\end{align}
Therefore, it is legitimate to use a new basis of 
$(N^{\prime c}_1,N^{\prime c}_2,N^{\prime c}_3,S^{\prime}_-)^T\equiv
V^\dag(N^c_1,N^c_2,N^c_3,S_-)^T$.  In this basis, it is
straightforward to find that $S'_\pm$ only couple to $N^{\prime c}_3$. Then,
the superpotential~\eqref{eq:W_hyb} is written as 
\begin{align}
  W_{\rm hyb}
  =
  \tilde{\lambda} S'_+S'_- N^{\prime c}_3 +
  \frac{1}{2}M'_{ij}N^{\prime c}_i N^{\prime c}_j\,,
  \label{eq:W_lambda_N_prime}
\end{align}
where $M^\prime\equiv W^T M W$ is a $3\times 3$ matrix of order $|M_{ij}|$.
Therefore, $\phi\equiv \sqrt{2}\mathrm{Re}\,\tilde{N}^{\prime c}_3$
becomes the inflaton field~\cite{Gunji:2019wtk,Gunji:2021zit}.

The parameter $\alpha$ can be taken to various values and different
predictions for $n_s$ and $r$ are obtained
accordingly, which is intensively analyzed in
Ref.~\cite{Gunji:2021zit}. In the current study, we take
\begin{align}
  \alpha&=2/3\,,~~~
  \tilde{\lambda} = 0.959\times10^{-3}\,, \quad
  \xi = (0.604\times10^{16}\,\mathrm{GeV})^2\,,
  \label{eq:lam_xi}
\end{align}
and $q=y=1$.  Then the best-fit value of observed spectral index
$n_s$, along with the tensor-to-scalar ratio $r\simeq6\times10^{-4}$,
is obtained at the last 60 $e$-folds of the subcritical regime of the
hybrid inflation~\cite{Gunji:2021zit}. With the parameters, the
condition~\eqref{eq:assumption} gives
\begin{align}
  |M'_{ij}|\ll 10^{13}\,{\rm GeV}\,,
  \label{eq:cond_Mij}
\end{align}
where $\expval{S'_+} =
(\xi/(2/3) q(1+\tilde{\xi}))^{1/2}$ 
and $\tilde{\xi}\equiv \xi/2q$
has been used, which will be
shown in the next subsection.  In the following subsections, we will
quantitatively examine the condition~\eqref{eq:cond_Mij} to keep the
inflaton dynamics unchanged. To keep the readable analytic
expressions, we leave $\tilde{\lambda}$, $\xi$, $q$ and $y$ as they
are.

\subsection{The scalar potential}
\label{sec:V_scalar}

The scalar potential is given by~\cite{Gunji:2021zit},
\begin{align}
  V
  &=
  V_F+V_D\,,\\
  V_F
  &=
  \Bigl(
    -\frac{\Phi}{3}
  \Bigr)^{-1}
  \frac{3}{2}
  \bigl[
    \delta^{I\bar{J}}W_I\overline{W}_{\bar{J}}
    +\frac{1}{\Delta}
    \bigl|\delta^{I\bar{J}}W_I\Phi_{\bar{J}}-2 W\bigr|^2
    +
    \frac{2}{\Phi}|W|^2
  \bigr]\,,\\
  V_D
  &=
  \frac{y^2}{2}
  \Bigl[
    \Bigl(
      -\frac{\Phi}{3}
    \Bigr)^{-1}
    \frac{2q}{3}
    \bigl(
      |S'_+|^2-|S'_-|^2
    \bigr)
    -\xi
  \Bigr]^2\,,
\end{align}
where $V_F$ and $V_D$ are $F$ and $D$-term potentials, respectively.
Here $W_I \equiv \partial W/\partial z^I\,,\, \Phi_I \equiv
\partial\Phi/\partial z^I\,$, and $\Delta \equiv
\Phi-\delta^{I\bar{J}}\Phi_I\Phi_{\bar{J}}$, and $z^I$ shows the
scalar component of $Z^I$.  In the $D$-term potential, $y$ and
$\xi\,(>0)$ are the gauge coupling and the Fayet-Iliopoulos (FI) term
related to the local U(1).  Due to the FI-term, $S'_+$ acquires a VEV
as $\langle S'_+\rangle = (\xi/(2/3) q(1+\tilde{\xi}))^{1/2}$ at the
global minimum.\footnote{From Eq.~\eqref{eq:lam_xi}, it is clear that
$\expval*{S_+}\ll 1$. } Then, $s\equiv \sqrt{2}|S'_+|$ is identified
with the waterfall field.

During inflation, we expect that the fields except for the inflaton
and waterfall fields are stabilized at the origin. Thus the
scalar potential during inflation is given as 
\begin{align}
  V(\phi,\,s)
  &\equiv
  V|_{\sqrt{2}\mathrm{Re}\tilde{N}^{\prime c}_3=
    \phi,\,\sqrt{2}|S'_+|=s,\,\text{the others}=0}\nonumber\\
  &=
  \Bigl(
    -\frac{\Phi(\phi,\,s)}{3}
  \Bigr)^{-1}
  \frac{3\phi^2}{4}
  \Bigl[
  \frac{\tilde{\lambda}^2}{2}
  s^2
  +
  \varDelta M^2(\phi,\,s)
  \Bigr]\nonumber\\
  &\quad
  +
  \frac{y^2}{8}
  \Bigl[
    \Bigl(
      -\frac{\Phi(\phi,\,s)}{3}
    \Bigr)^{-1}
    \frac{2q}{3} s^2
    -2\xi
  \Bigr]^2\,,
\label{eq:Vphis}
\end{align}
where $\Phi(\phi,\,s) \equiv -3+(s^2+\phi^2)/2$. $\Delta M^2(\phi,s)$ is
the term that originates in the Majorana mass term, which is given by
\begin{align}
  \varDelta M^2(\phi,\,s)
  &\equiv
  \sum_{i=1}^3|M^\prime_{i3}|^2
  -
  |M^\prime_{33}|^2
    \frac{\phi^2}{12}
    \Bigl(
      -\frac{\Phi(\phi,\,s)}{3}
    \Bigr)^{-1}\,.
\end{align}

\subsection{Critical point}

As in the
literature~\cite{Buchmuller:2014rfa,Buchmuller:2014dda,Ishiwata:2018dxg,Gunji:2019wtk,Gunji:2021zit},
we focus on the dynamics of the subcritical regime of the hybrid
inflation. First of all, the Majorana mass term shifts the critical
point value of the hybrid inflation. The critical point value is
determined by the mass squared of the waterfall field, which is given
by
\begin{align}
  m_+^2
  =
  m_{+,0}^2
  +
  \varDelta m_{+}^2\,,
\end{align}
where
\begin{align}
  m_{+,0}^2
  &\equiv
  q y^2 \xi\bigl(\Psi(\phi)-1\bigr)
  \,,\\
  \varDelta m_+^2
  &\equiv
  qy^2\xi
  \Psi(\phi)
  \frac{1}{3\tilde{\lambda}^2}
  \Bigl(
    -\frac{\Phi_0}{3}
  \Bigr)^{-1}\nonumber\\
    &\quad\times
  \Bigl[
    \varDelta M^2(\phi,0)
    +
    |M^\prime_{33}|^2
    \frac{\phi^2}{12}
    \Bigl(
      -\frac{\Phi_0}{3}
    \Bigr)^{-1}
  \Bigr]\,,\\
  \Psi(\phi)
  &\equiv
  \frac{\tilde{\lambda}^{2}}{2(2/3)^2qy^2\xi}\phi^2\,,
  \quad
  \Phi_0\equiv \Phi(\phi,0)
  \,.
\end{align}
Here we have given the mass in a canonically normalized basis in
accordance with Ref.~\cite{Gunji:2021zit}. Consequently, the
critical point value $\phi_c$ is obtained by
\begin{align}
  \phi_c^2
  &=
  \phi_{c,0}^2
  (1-\delta)
  +\mathcal{O}(M_{i3}^{\prime4})\,,
\end{align}
where $\phi_{c,0}^2=
{2(2/3)^2qy^2\xi}/{\tilde{\lambda}^{2}}$~\cite{Gunji:2021zit} and
\begin{align}
  \delta
  &\equiv
  \frac{2(2/3)^2}{\tilde{\lambda}^{2}}
  \frac{\varDelta m_+^2(\phi_{c,0})}{\phi_{c,0}^2}\,.
\end{align}
Imposing $|\delta|\ll1$, the upper bound on $|M^\prime_{i3}|$ is
obtained as
\begin{align}
  \sum_{i=1}^2
  |M^\prime_{i3}|^2
  &\ll
  3\tilde{\lambda}^{2}-4qy^2\xi/9
  \simeq
  3\times10^{-7}
  \sim(1\times10^{15}\, \mathrm{GeV})^2\,,\label{eq:Mi3_phic_a=2ov3_c=0}\\
  |M^\prime_{33}|^2
  &\ll
  \frac
    {(3\tilde{\lambda}^2-4qy^2\xi/9)^2}
    {3\tilde{\lambda}^2(3\tilde{\lambda}^2-8qy^2\xi/9)}
  \simeq
  4\times10^{-8}
  \sim (5\times10^{14}\,\mathrm{GeV})^2\,.
  \label{eq:M33_phic_a=2ov3_c=0}
\end{align}
Those are weaker bounds compared to Eq.~\eqref{eq:cond_Mij}. Therefore,
the critical point value is merely affected by the Majorana mass term.

\subsection{Subcritical regime}

Below the critical point, the waterfall field grows due to the
tachyonic instability~\cite{Asaka:2001ez,Buchmuller:2014rfa} and soon
relaxes to the local minimum $s_{\mathrm{min}}$ of the classical
path.\footnote{ We have confirmed that the one-loop potential
dominates over the Majorana mass terms near the critical point under
the condition~\eqref{eq:cond_Mij}. Thus, the dynamics around the
critical point is the same as one studied in
Ref.~\cite{Gunji:2021zit}.} $s_{\mathrm{min}}$ is obtained by
$\partial V(\phi,s)/\partial s^2 = 0$ as
\begin{align}
  s_{\mathrm{min}}^2
  &=
  \frac{3\xi}{q}
  \frac
  {(-\Phi_0/3)}
  {1+\tilde{\xi}(1-\Psi(\phi))}\nonumber\\
  &\quad\times
  \Biggl[
    1-
    \Psi(\phi)
    \Biggl\{
      1-
      \frac
      {
        \tilde{\xi}\bigl[1-\Psi(\phi)\bigr](\Delta_+-\Delta_-)
        -\Delta_-
      }
      {
        1+\tilde{\xi}
        \bigl[
          1-\Psi(\phi)(1+\Delta_+)
        \bigr]
      }
    \Biggr\}
  \Biggr]
    \,,\label{eq:smin_a=2ov3}
\end{align}
where
\begin{align}
  \Delta_\pm
  &\equiv
  \Bigl(
    -\frac{\Phi_0}{3}
  \Bigr)^{-1}
  \frac{1}{3\tilde{\lambda}^2}
  \Bigl[
    \varDelta M^2(\phi,0)
    \pm
    |M^\prime_{33}|^2
    \frac{\phi^2}{12}
    \Bigl(
      -\frac{\Phi_0}{3}
    \Bigr)^{-1}
  \Bigr]\,.
\end{align}
Consequently the potential in the subcritical regime can be expressed
by a single field effective potential
\begin{align}
  V_{\mathrm{sub}}(\phi)
  &\equiv
  V(\phi,\,s_{\mathrm{min}}(\phi))\nonumber\\
  &=
  y^2\xi^2\Psi(\phi)
  \Biggl[
    1-\frac{\Psi(\phi)}{2}
    +
    \frac
    { 1+\tilde{\xi} \{1-\Psi(\phi)\} }
    { 3\tilde{\lambda}^2\tilde{\xi} }
    \Bigl(
      -\frac{\Phi_0}{3}
    \Bigr)^{-1}\nonumber\\
  &\qquad \qquad\times
    \Bigl[
      \varDelta M^2(\phi,0)
      -
      |M^\prime_{33}|^2
      \frac{\phi^2}{12}
       \Bigl(
        -\frac{\Phi_0}{3}
      \Bigr)^{-1}
      \tilde{\xi}
      \{1-\Psi(\phi)\}
    \Bigr]
  \Biggr]
  +
  \mathcal{O}(M_{i3}^{\prime 4})
  \nonumber \\
  &=
  y^2\xi^2\Psi(\phi)
  \Biggl[
    1-\frac{\Psi(\phi)}{2}
    +
    \frac
  {1}{3\tilde{\lambda}^2\tilde{\xi}}
  \Bigl(
    -\frac{\Phi_0}{3}
  \Bigr)^{-1}
    \varDelta M^2(\phi,0)
  \Biggr]
  +
  \mathcal{O}(M_{i3}^{\prime 4},\xi^3)
  \,.\label{eq:Vsub_a=2ov3}
\end{align}
In the last step, we have used $\xi\ll 1$.
Therefore,  $V_{\mathrm{sub}}$ reduces to the inflaton potential in
Ref.~\cite{Gunji:2021zit} when 
\begin{align}
  |\varDelta M^2(\phi,0)|
  &\ll
  3\tilde{\lambda}^2\tilde{\xi}
  \Bigl(
    -\frac{\Phi_0}{3}
  \Bigr)
  \Bigl[
  1-\frac{\Psi(\phi)}{2}
  \Bigr]
  \,,\label{eq:Vsub1st_vs_2nd_a=2ov3}
\end{align}
is satisfied. This condition gives rise to upper bounds on
$|M^\prime_{i3}|$ as
\begin{align}
  \sum_{i=1}^2
  |M^\prime_{i3}|^2
  &\ll
  3\tilde{\xi}\tilde{\lambda}^2
  ( 1+\phi_{c,0}^2/6)
  /2
  \sim(2\times10^{12}\,\mathrm{GeV})^2\,,
  \label{eq:Mi3_subcrit_a=2ov3_c=0}
  \\
  |M^\prime_{33}|^2
  &\ll
  2\tilde{\xi}\tilde{\lambda}^2
  {(1-\phi_{c,0}^2/6)^2}/{|4-\phi_{c,0}^2|}
  \sim(9\times10^{11}\,\mathrm{GeV})^2\,.
  \label{eq:M33_subcrit_a=2ov3_c=0}
\end{align}
Those are a bit tighter than the condition~\eqref{eq:cond_Mij}.  When
the above conditions are satisfied, $s_{\mathrm{min}}$ is also in
approximate agreement with the local minimum value of the waterfall
field in Ref.~\cite{Gunji:2021zit},
\begin{align}
  s_{\mathrm{min}}^2
  &=
  \frac{3\xi}{q}
  \Bigl(
    -\frac{\Phi_0}{3}
  \Bigr)
  \Bigl[
    1-\Psi(\phi)
  \Bigr]+\mathcal{O}(\xi\Delta_-)\,.
  \label{eq:smin_approx}
\end{align}

To summarize the inflaton dynamics is not affected if 
\begin{align}
  |M^\prime_{i3}|\ll10^{12}\,\mathrm{GeV}\,,
  \label{eq:upper_bound_on_M_i3}
\end{align}
is satisfied.\footnote{We have checked that the stability of the
  $\tilde{L}_iH_u$ direction, pointed out by
  Ref.~\cite{Nakayama:2016gvg}, is guaranteed if $|M'_{ij}|\lesssim
  10^{18}\,{\rm GeV}$.  } In the next subsection, we will confirm the
condition numerically.

\subsection{Dynamics of scalar fields: numerical study}
\label{sec:inf_num}

\begin{figure}[tbp]
  \begin{minipage}[t]{0.5\hsize}
  \centering
  \includegraphics[keepaspectratio,width=7.8cm]{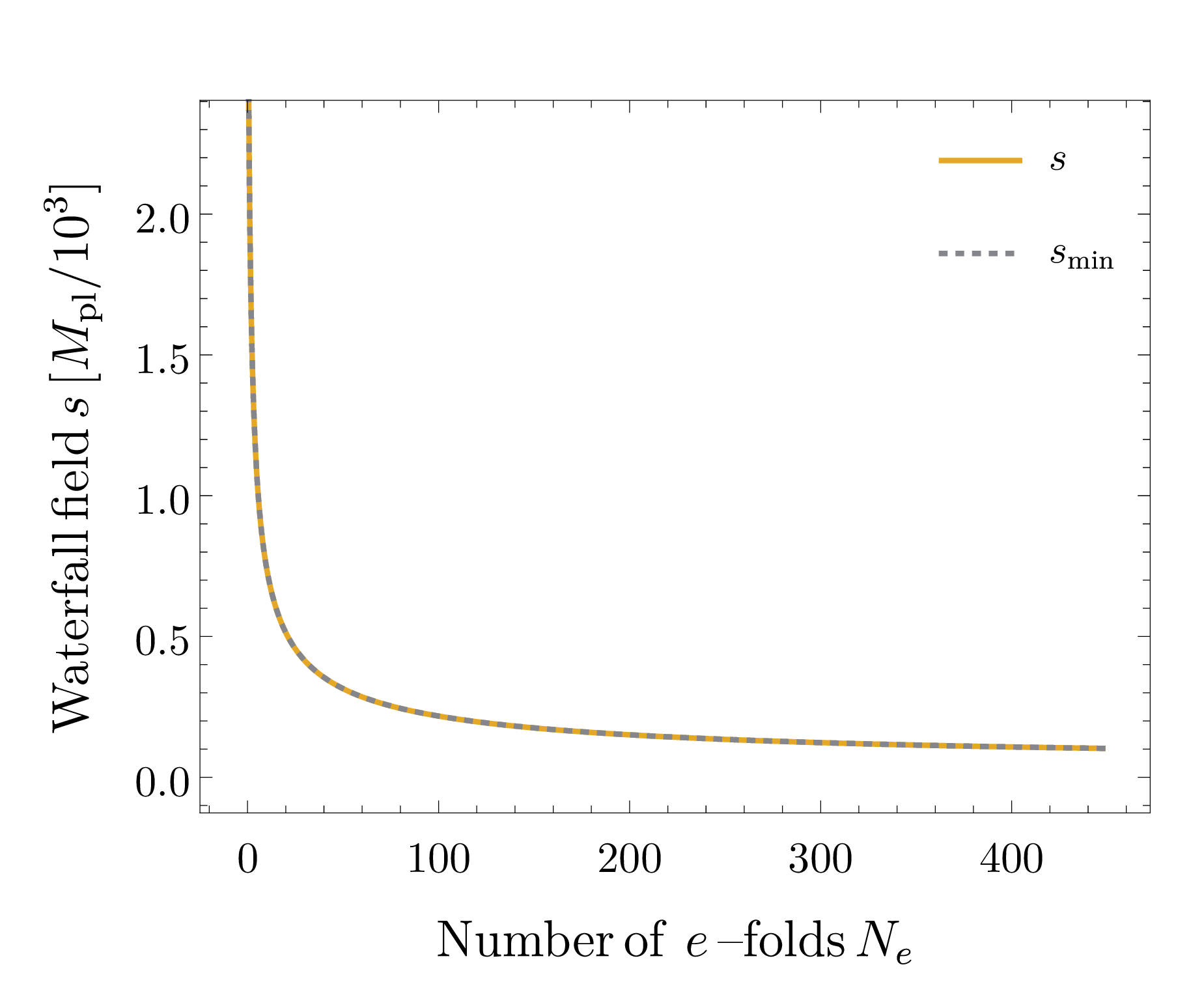}
  \end{minipage}
  \begin{minipage}[t]{0.5\hsize}
  \centering
  \includegraphics[keepaspectratio,width=7.8cm]{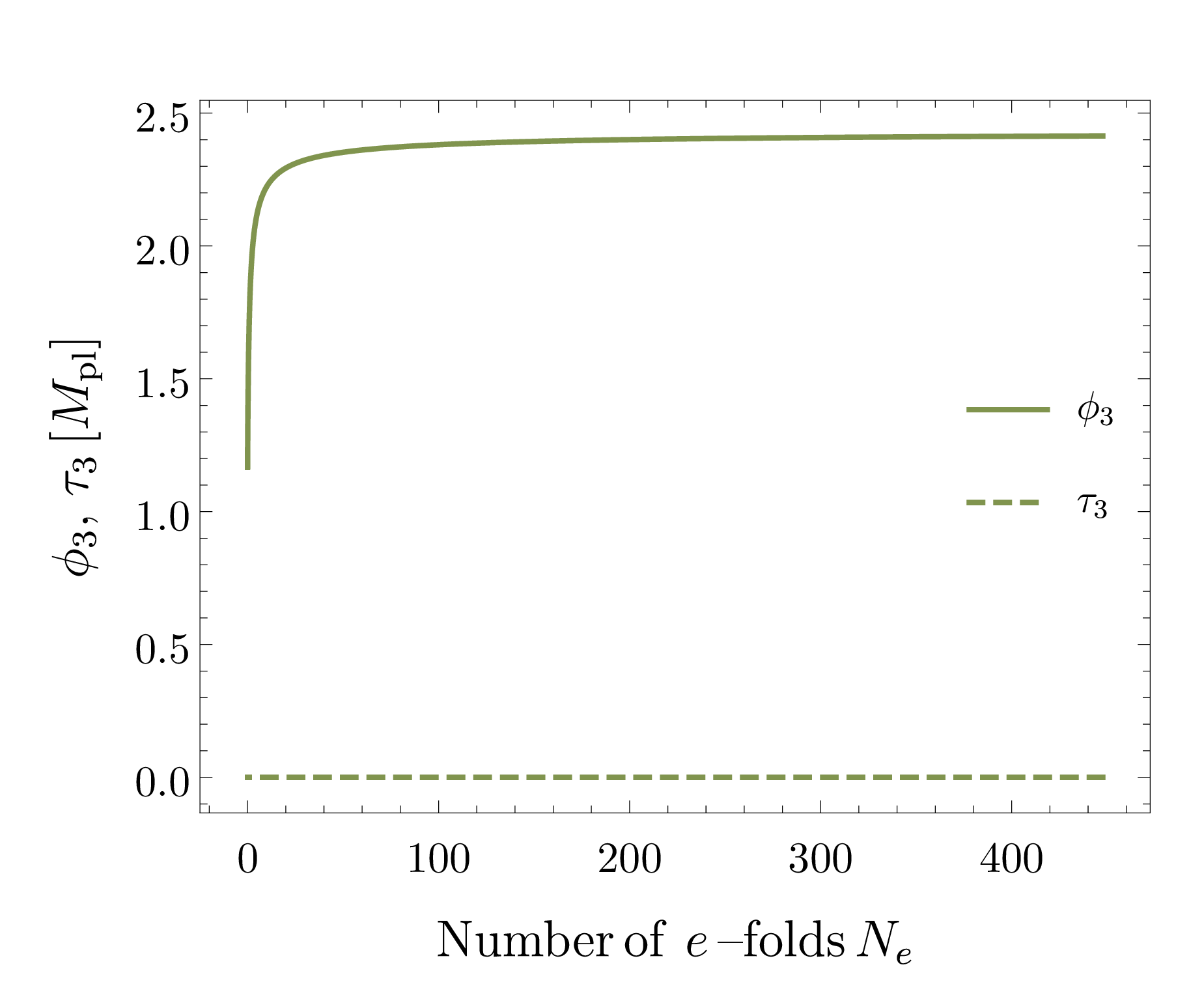}
  \end{minipage}
  \begin{minipage}[t]{0.5\hsize}
  \centering
  \includegraphics[keepaspectratio,width=7.8cm]{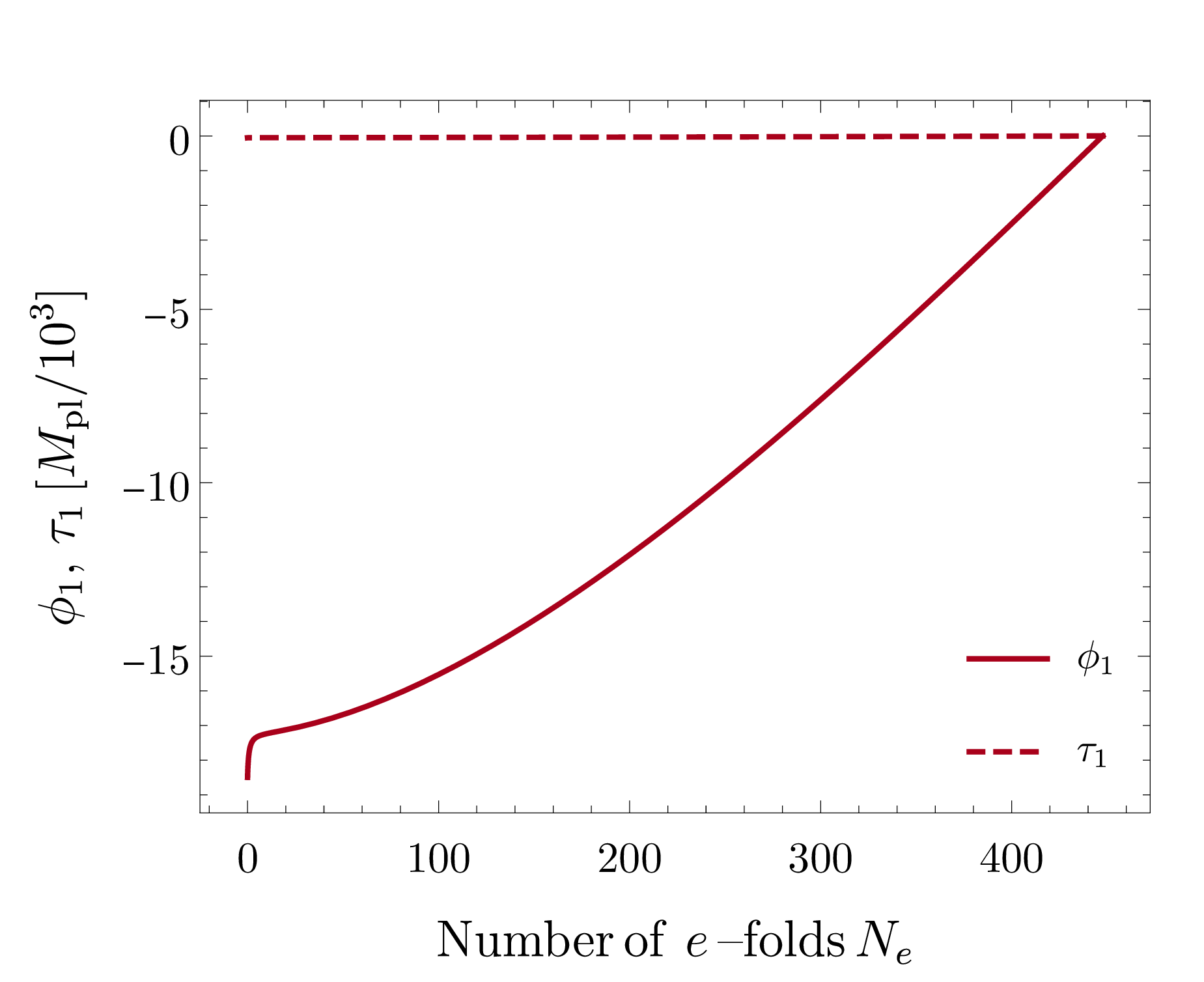}
  \end{minipage}
  \begin{minipage}[t]{0.5\hsize}
  \centering
  \includegraphics[keepaspectratio,width=7.8cm]{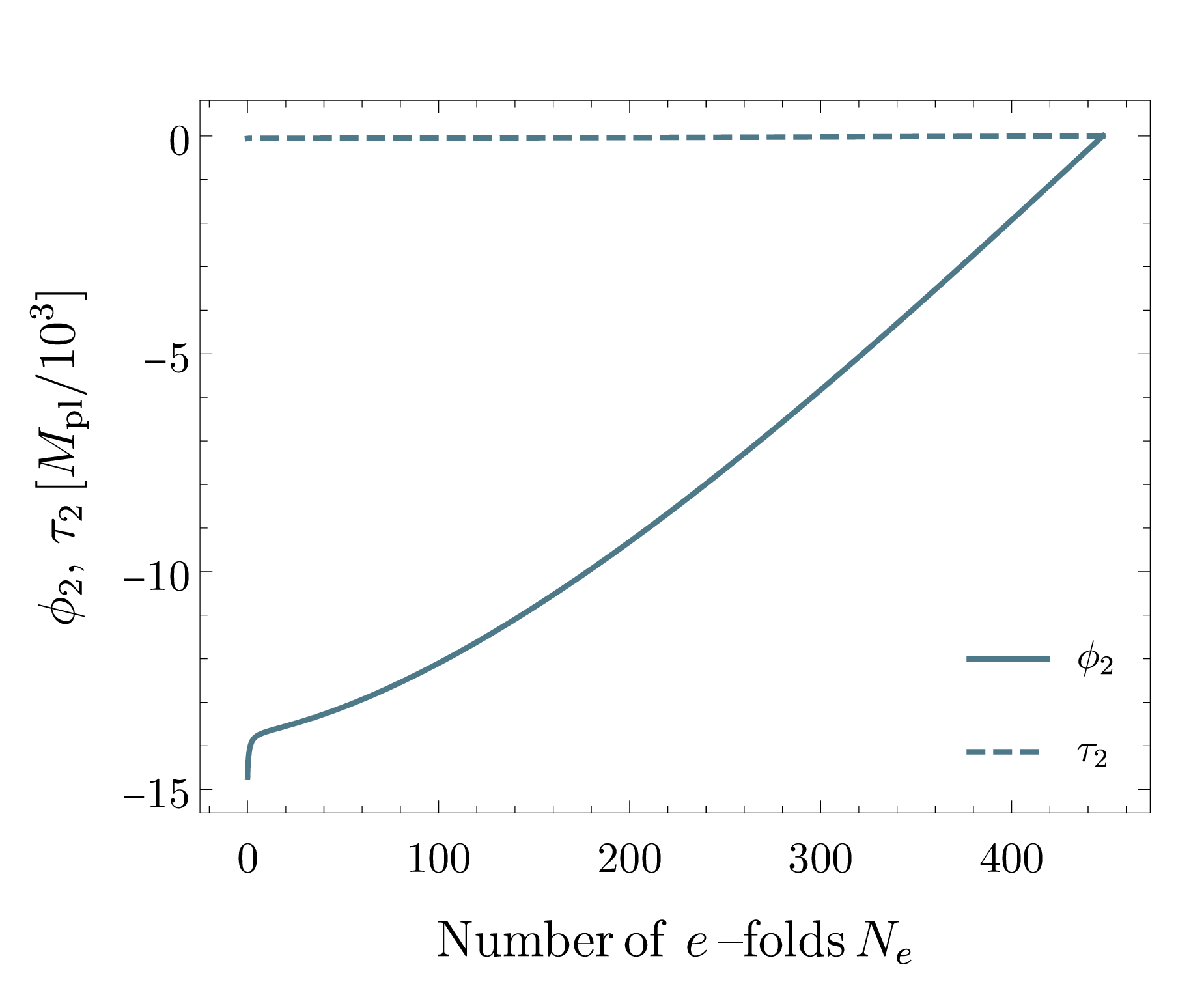}
  \end{minipage}
  \caption{The time evolution of $s$, $\phi_i$, and $\tau_i$
    $(i=1 \text{--} 3)$ as function of $e$-folds before the end of inflation.
    The parameters are given in Eq.~\eqref{eq:bm1} and
    $\hat{\Lambda}=10^{10}\,\mathrm{GeV}$.
    (Top-left) The waterfall field value $s$ (solid) given by
    the numerical solution and that of the analytical approximation
    $s_{\mathrm{min}}$ (dotted) are shown.  (Top-right) The
    inflaton $\phi_3$ (solid) and $\tau_3$ (dashed) are given.
    (Bottom) $\phi_i$ (solid) and $\tau_i$ (dashed) for $i=1,2$ are
    shown.  }
  \label{Fig:varphi_vs_phi}
\end{figure}

The upper bound~\eqref{eq:upper_bound_on_M_i3} is obtained under the
assumption that the fields except for the inflaton and waterfall
fields are stabilized at the origin during inflation.  However, we
need to investigate the validity of this assumption since the values
of $\tilde{N}^{\prime c}_{1,2}$ and $\mathrm{Im}\,\tilde{N}^{\prime
  c}_3$ may grow to affect the dynamics of inflation depending on the
value of the Majorana masses.  We will check the stability of
$\tilde{N}^{\prime c}_{1,2}$ and $\mathrm{Im}\,\tilde{N}^{\prime c}_3$
by solving the equations of motion of $\tilde{N}^{\prime c}_i$ and
$S'_+$ numerically and examine the
condition~\eqref{eq:upper_bound_on_M_i3} more quantitatively.

To this end, we define the relevant scalar fields as
$\{\phi_i,\,\tau_i,\,s\}\equiv\varphi^A$ where\footnote{Namely,
$\phi_3=\phi$ is the inflaton field in the notation of this
subsection.}
\begin{align}
  \phi_i\equiv\sqrt{2}\mathrm{Re}\,\tilde{N}^{\prime c}_i\,,\,
  \tau_i\equiv\sqrt{2}\mathrm{Im}\,\tilde{N}^{\prime c}_i\,.\,
\end{align}
Then, the metric in terms of the field space of $\varphi^A$ is given
by
\begin{align}
  G_{AB}=\delta_{AB}\left\{
  \begin{array}{ll}
    K^{\rm m}_{N_i\bar{N}_i} & {\rm for}~\varphi^A=\phi_i\,,\tau_i \\
    K^{\rm m}_{S^+\bar{S}^+} & {\rm for}~\varphi^A=s
    \end{array}
  \right.\,,
\end{align}
where $K_{Z^I\bar{Z}^J}=\pdv*{K}{z^I}{\bar{z}^J}$.  The equations of
motion of $\varphi^A$ are
\begin{align}
  \ddot{\varphi}^A+3H\dot{\varphi}^A
  +G^{AB}\frac{\partial V}{\partial \varphi^B}
  +\Gamma^A_{BC}\dot{\varphi}^B\dot{\varphi}^C=0\,.
\end{align}
Here, dot denotes the time derivative and $H$ is the Hubble parameter
that depends on $\varphi^A$ and $\dot{\varphi}^A$. $G^{AB}$ is the
inverse of $G_{AB}$ and $\Gamma^A_{BC}$ is the connection defined by
\begin{align}
  \Gamma^A_{BC}
  \equiv
  \frac{1}{2}G^{AD}
  (G_{DB,C}+G_{DC,B}-G_{BC,D})\,.
\end{align}
For the numerical analysis, we take a benchmark point from the
allowed region of $\tau$ and $\hat{g}$, which is given by
\begin{align}
  \begin{split}
    \mathrm{Re}\,\tau &= 8.88\times10^{-4}\,,\quad
  \mathrm{Im}\,\tau = 1.08\,,\\
    g &=  4.56\,,\quad
  \phi_g= -170^\circ\,.
  \end{split}
  \label{eq:bm1}
\end{align}
With the parameters the corresponding neutrino mixing parameters are 
\begin{align}
  \begin{split}
    \sin^2\theta_{12}&=0.305\,,\quad
    \sin^2\theta_{13}=0.0224\,,\quad
    \sin^2\theta_{23}=0.572\,,\\
    \delta_{\mathrm{CP}}&=-160^\circ\,,\quad
    \alpha_{21}=-163^\circ\,,
  \end{split}
\end{align}
and $|M^\prime_{ij}|$ is obtained as
\begin{align}
  |M^\prime_{ij}|
  &=
  \begin{pmatrix}
    5.59 &	3.56 &	9.75 \\
    3.56 &	6.52 &	6.57 \\
    9.75 &	6.57 &	3.53 
  \end{pmatrix}
  \times 10^{9}
  \,\mathrm{GeV}
  \,
  \Bigl(
  \frac{\hat{\Lambda}}{10^{10}\,\mathrm{GeV}}
  \Bigr)\,.
\end{align}
Since we are considering the subcritical regime, the initial values of
$\phi$ and $s$ are set to $\phi^2_{\mathrm{init}}\equiv
0.98\phi_{c,0}^2$ and $s_{\mathrm{init}}\equiv
s_{\mathrm{min}}(\phi_{\mathrm{init}})$, respectively, at the time
$t=0$, while the other initial field values are set to zero.  We have
confirmed that similar trajectories of the scalar fields are obtained
for slightly different the initial values of the inflaton and
waterfall fields.\footnote{The waterfall field value is
much smaller than the inflaton one during inflation and the trajectory
is almost straight along the inflaton field direction.  Therefore, the
trajectory of $\phi_3$-$s$ system has no additional impact on the
curvature perturbation, which is already studied in
Ref.~\cite{Gunji:2021zit}.}

\begin{figure}[tbp]
  \begin{minipage}[t]{0.5\hsize}
  \centering
  \includegraphics[keepaspectratio,width=7.8cm]{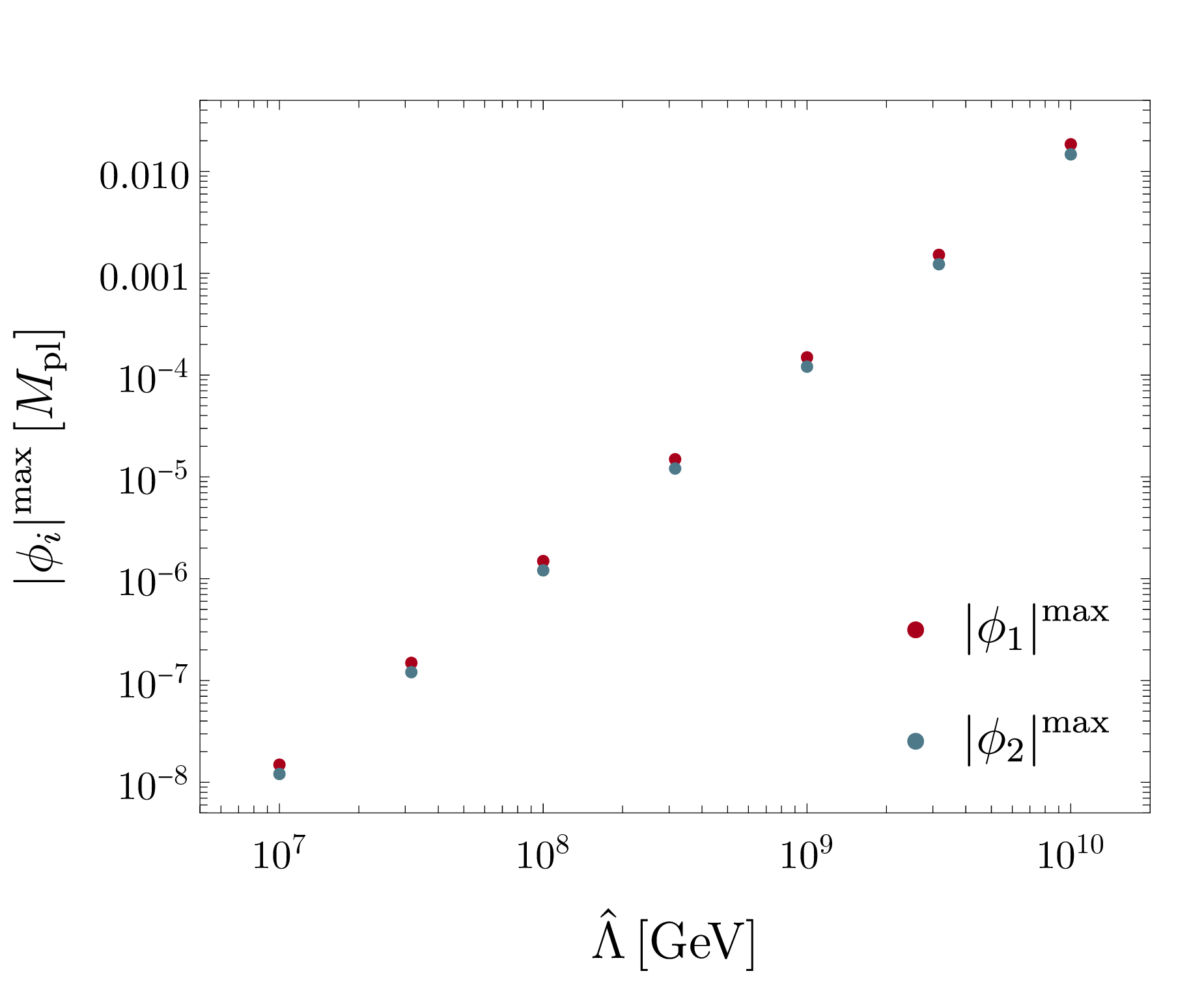}
  \end{minipage}
  \begin{minipage}[t]{0.5\hsize}
  \centering
  \includegraphics[keepaspectratio,width=7.8cm]{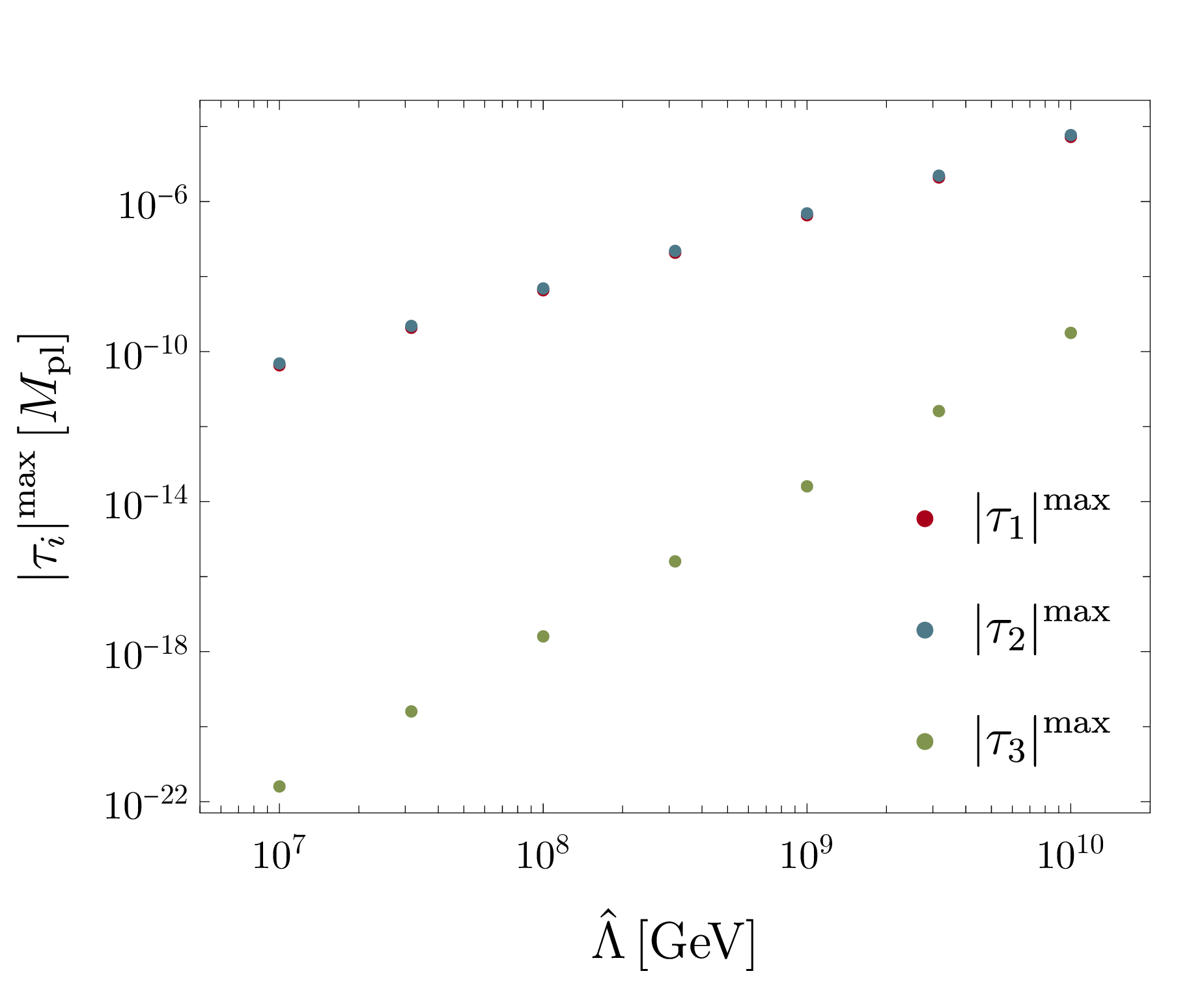}
  \end{minipage}
  \caption{ The maximum values of $\phi_i\,\,(i=1,2)$ (left) and
    $\tau_i\,\,(i=1 \text{--} 3)$ (right) during inflation as
    functions of $\Lambda$. Red, blue, and green points correspond to
    $i=1$, $2$, and $3$, respectively.  }
  \label{Fig:varphi_vs_Lambda}
\end{figure}

Fig.~\ref{Fig:varphi_vs_phi} shows the time evolution of $\varphi^A$
before the end of inflation as a function of the $e$-folds $N_e$ defined
by
\begin{align}
  N_e(t) = \int^{t_{\rm end}}_tdt' H\,,
\end{align}
where $t_{\rm end}$ is the time at the end of inflation. In the plot, $\hat{\Lambda}=10^{10}\,\mathrm{GeV}$ is taken to satisfy the
condition \eqref{eq:upper_bound_on_M_i3} at a percent level.
We confirmed that the trajectory of the waterfall field well agrees
with the local minimum $s_{\rm min}$, given in
Eq.~\eqref{eq:smin_approx}. In addition, we found that the $\tau_i$
$(i=1 \text{--} 3)$ merely move during the inflation. On the other hand,
$\phi_1$ and $\phi_2$ grow as large as $\order{\expval{s}}$ as seen in
the figure.\footnote{$\phi_i$ ($i=1,2$) are kicked by the term
proportional to $\sum_{j}\Re (M_{ij}^*M_{j3})\phi_i\phi_3$. The
direction that $\phi_i$ is heading for depends on the sign of
$\sum_{j}\Re (M_{ij}^*M_{j3})$.} They are, however, still subdominant
components in the scalar potential and they do not have any impact on
the waterfall-inflaton dynamics. For smaller value of $\hat{\Lambda}$
we found that the field values of $\phi_1$ and $\phi_2$ reduce almost
linearly, which is summarized in Fig.~\ref{Fig:varphi_vs_Lambda}.
Having confirmed the condition \eqref{eq:upper_bound_on_M_i3}, we
derived more quantitative bound,
\begin{align}
  \hat{\Lambda}\lesssim 10^{10}\,\mathrm{GeV}\,,
\end{align}
or equivalently
\begin{align}
  |M^\prime_{i3}|\lesssim 10^{10}\,\mathrm{GeV}\,.
\end{align}

\begin{figure}[tbp]
  \begin{minipage}[t]{0.5\hsize}
  \centering
  \includegraphics[keepaspectratio,width=7.8cm]{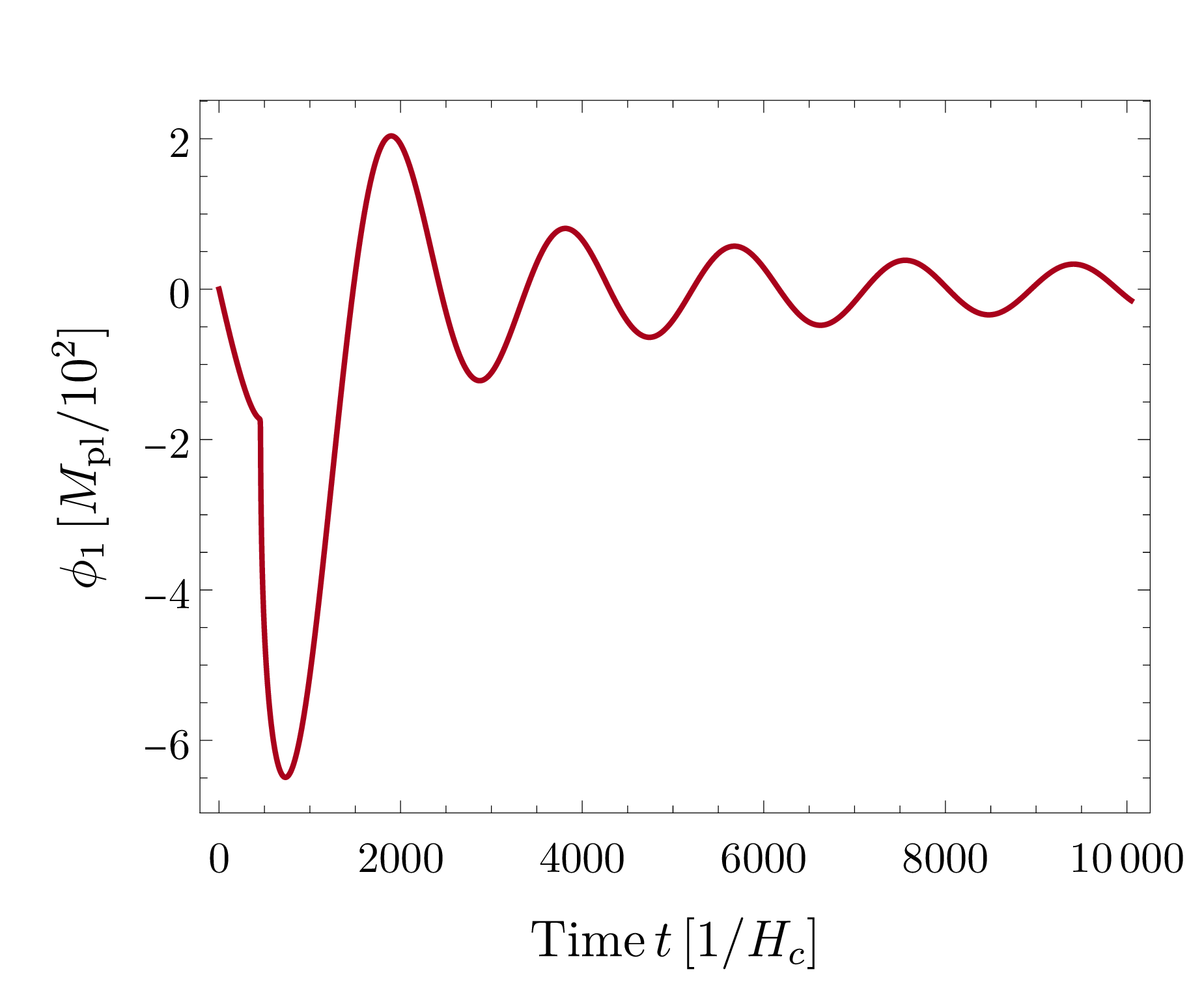}
  \end{minipage}
  \begin{minipage}[t]{0.5\hsize}
  \centering
    \includegraphics[keepaspectratio,width=7.8cm]{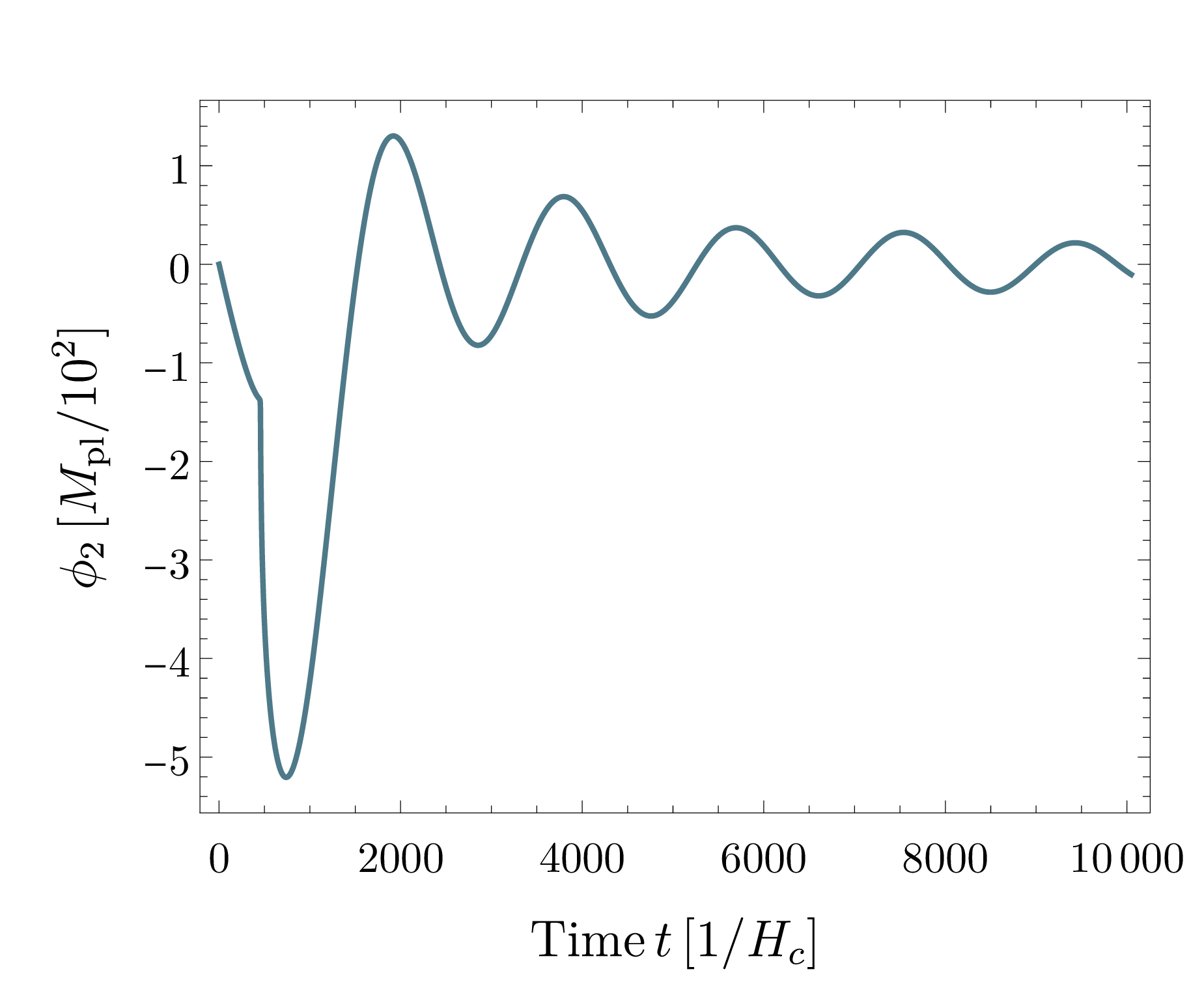}
  \end{minipage}
  \caption{Time evolution of $\phi_1$ and $\phi_2$ as function of the
    cosmic time $t$. Here $H_c\equiv H|_{\phi=\phi_c}\simeq
    6.1\times10^{12}\,\mathrm{GeV}$. The parameters are the same as
    Fig.~\ref{Fig:varphi_vs_phi}. }
  \label{Fig:phi1_2}
\end{figure}

In the bottom panels of Fig.~\ref{Fig:varphi_vs_phi}, the field
growths of $\phi_1$ and $\phi_2$ get
accelerated. This is because the Hubble parameter begins to decrease
after the end of inflation since the inflaton field oscillates around
the minimum to behave as a matter component. After that, the $\phi_1$
and $\phi_2$ start to oscillate with an angular frequency
$\order{|M_{ij}|}$. We checked this behavior numerically, which is
shown in Fig.~\ref{Fig:phi1_2}. Here we have assumed that the inflaton
field keeps the coherent oscillation. Realistically the inflaton field
decays to leptons and Higgsinos or sleptons and Higgses and it reheats
the universe. Then the radiation domination follows. Although $\phi_1$
and $\phi_2$ continue the coherent oscillation, we expect that they do
not become the dominant component of the universe because their energy
density is highly suppressed. In that case, non-thermal leptogenesis
by the inflaton field works and it is expected to provide a sufficient
number of lepton
asymmetry~\cite{Fukugita:1986hr,Nakayama:2016gvg,Bjorkeroth:2016qsk,Murayama:1992ua,Murayama:1993xu,Murayama:1993em,Hamaguchi:2001gw,Ellis:2003sq,Antusch:2004hd,Antusch:2009ty,Kadota:2005mt,Nakayama:2013nya,Gunji:2019wtk}. The
details depend on the model parameters and they are beyond the scope
of our current study.  We leave it for our future study.

\section{Conclusion}
\label{sec:conclusion}

We consider a supergravity model that has the modular $A_4$
symmetry. This model accommodates the MSSM augmented by three
right-handed neutrino fields that have the Majorana
masses. Additionally, two fields that are charged under a gauged U(1)
are introduced. They couple to the right-handed neutrinos via the
Yukawa interaction and consequently one of the scalar components plays
a role of the waterfall field during inflation and acquires a VEV at
the global minimum. With the extension, the pattern of the light
neutrino mass, generated by the seesaw mechanism, changes and one of
the light neutrinos becomes massless. On top of that, the modular
$A_4$ symmetry restricts the mixing pattern. Comparing with the
current observations regarding the neutrino mixings, we have found
that only the IH case is allowed. The predicted Majorana phase is
around 
$\pm 35^\circ$,
the Dirac phase is 
$[-110^\circ,-90^\circ],~[90^\circ,110^\circ]$
and $0.42
\lesssim \sin^2 \theta_{23} \lesssim 0.43$, or the Majorana phase is
$[-180^\circ,-120^\circ],~[120^\circ,180^\circ]$, 
the Dirac phase takes values in
$[-180^\circ,180^\circ]$
and $0.42 \lesssim \sin^2 \theta_{23} \lesssim 0.62$. The
effective neutrino mass, which determines the decay rate of the
neutrinoless double beta decay, is found to be around 14--28~meV
and 45--47~meV. Such a relatively large effective mass of
$\mathcal{O}(10)$~meV will be explored in future experiments.

The supergravity model we consider is based on a hybrid inflation
model that has the subcritical inflation regime. Namely, inflation
continues below the critical point. In the present model, the three
right-handed sneutrinos are the candidates for the inflaton field.
Compared to the inflation model studied in Ref.~\cite{Gunji:2021zit},
the existence of the Majorana mass terms is a crucial difference, which
may affect the inflation dynamics. Imposing the Majorana mass not to
affect the inflation dynamics, we have revealed that only one of the
right-handed sneutrinos turns out to couple to the waterfall field,
then we have derived the upper bound for the Majorana mass scale
analytically. We have confirmed the results numerically by solving the
equations of motion for scalar fields. It is found
the other right-handed sneutrinos grow as large as the VEV of the
waterfall field. They are, however, negligible in the total energy
density during the inflation if the Majorana mass scale is smaller
than $\order{10^{10}}~{\rm GeV}$.  Though their field values are much
suppressed during inflation, the scalar fields continue to oscillate
after inflation, which may affect the subsequent thermal history. For
instance, they may contribute to the generation of the lepton
asymmetry. The details depend on the parameter and we leave it for
the future work.

\section*{Acknowledgments}

We thank Tatsuo Kobayashi for valuable discussion at the early stage
of this project and Takehiko Asaka for useful comments on the
preprint. This work is supported by JST SPRING, Grant No. JPMJSP2135
(YG), JSPS KAKENHI Grant No. JP18H05542, JP20H01894, JSPS Core-to-Core
Program Grant No. JPJSCCA20200002 (KI), and JSPS KAKENHI Grant
No. 20H01898 (TY).
\newpage
\bibliography{draft}

\providecommand{\href}[2]{#2}\begingroup\raggedright\begin{thebibliography}{100}

\bibitem{Akrami:2018odb}
{\scshape Planck} collaboration, \emph{{Planck 2018 results. X. Constraints on
  inflation}}, \href{https://doi.org/10.1051/0004-6361/201833887}{\emph{Astron.
  Astrophys.} {\bfseries 641} (2020) A10}
  [\href{https://arxiv.org/abs/1807.06211}{{\ttfamily 1807.06211}}].

\bibitem{Feruglio:2017spp}
F.~Feruglio, \emph{{Are neutrino masses modular forms?}},
  \href{https://arxiv.org/abs/1706.08749}{{\ttfamily 1706.08749}}.

\bibitem{Adamson:2013whj}
{\scshape MINOS} collaboration, \emph{{Measurement of Neutrino and Antineutrino
  Oscillations Using Beam and Atmospheric Data in MINOS}},
  \href{https://doi.org/10.1103/PhysRevLett.110.251801}{\emph{Phys. Rev. Lett.}
  {\bfseries 110} (2013) 251801}
  [\href{https://arxiv.org/abs/1304.6335}{{\ttfamily 1304.6335}}].

\bibitem{Adamson:2013ue}
{\scshape MINOS} collaboration, \emph{{Electron neutrino and antineutrino
  appearance in the full MINOS data sample}},
  \href{https://doi.org/10.1103/PhysRevLett.110.171801}{\emph{Phys. Rev. Lett.}
  {\bfseries 110} (2013) 171801}
  [\href{https://arxiv.org/abs/1301.4581}{{\ttfamily 1301.4581}}].

\bibitem{Abe:2017vif}
{\scshape T2K} collaboration, \emph{{Measurement of neutrino and antineutrino
  oscillations by the T2K experiment including a new additional sample of
  $\nu_e$ interactions at the far detector}},
  \href{https://doi.org/10.1103/PhysRevD.96.092006}{\emph{Phys. Rev. D}
  {\bfseries 96} (2017) 092006}
  [\href{https://arxiv.org/abs/1707.01048}{{\ttfamily 1707.01048}}].

\bibitem{Abe:2018wpn}
{\scshape T2K} collaboration, \emph{{Search for CP Violation in Neutrino and
  Antineutrino Oscillations by the T2K Experiment with $2.2\times10^{21}$
  Protons on Target}},
  \href{https://doi.org/10.1103/PhysRevLett.121.171802}{\emph{Phys. Rev. Lett.}
  {\bfseries 121} (2018) 171802}
  [\href{https://arxiv.org/abs/1807.07891}{{\ttfamily 1807.07891}}].

\bibitem{Adamson:2017gxd}
{\scshape NOvA} collaboration, \emph{{Constraints on Oscillation Parameters
  from $\nu_e$ Appearance and $\nu_\mu$ Disappearance in NOvA}},
  \href{https://doi.org/10.1103/PhysRevLett.118.231801}{\emph{Phys. Rev. Lett.}
  {\bfseries 118} (2017) 231801}
  [\href{https://arxiv.org/abs/1703.03328}{{\ttfamily 1703.03328}}].

\bibitem{NOvA:2018gge}
{\scshape NOvA} collaboration, \emph{{New constraints on oscillation parameters
  from $\nu_e$ appearance and $\nu_\mu$ disappearance in the NOvA experiment}},
  \href{https://doi.org/10.1103/PhysRevD.98.032012}{\emph{Phys. Rev. D}
  {\bfseries 98} (2018) 032012}
  [\href{https://arxiv.org/abs/1806.00096}{{\ttfamily 1806.00096}}].

\bibitem{Kobayashi:2018vbk}
T.~Kobayashi, K.~Tanaka and T.H.~Tatsuishi, \emph{{Neutrino mixing from finite
  modular groups}},
  \href{https://doi.org/10.1103/PhysRevD.98.016004}{\emph{Phys. Rev. D}
  {\bfseries 98} (2018) 016004}
  [\href{https://arxiv.org/abs/1803.10391}{{\ttfamily 1803.10391}}].

\bibitem{Criado:2018thu}
J.C.~Criado and F.~Feruglio, \emph{{Modular Invariance Faces Precision Neutrino
  Data}}, \href{https://doi.org/10.21468/SciPostPhys.5.5.042}{\emph{SciPost
  Phys.} {\bfseries 5} (2018) 042}
  [\href{https://arxiv.org/abs/1807.01125}{{\ttfamily 1807.01125}}].

\bibitem{Kobayashi:2018scp}
T.~Kobayashi, N.~Omoto, Y.~Shimizu, K.~Takagi, M.~Tanimoto and T.H.~Tatsuishi,
  \emph{{Modular A$_{4}$ invariance and neutrino mixing}},
  \href{https://doi.org/10.1007/JHEP11(2018)196}{\emph{JHEP} {\bfseries 11}
  (2018) 196} [\href{https://arxiv.org/abs/1808.03012}{{\ttfamily
  1808.03012}}].

\bibitem{Kobayashi:2018wkl}
T.~Kobayashi, Y.~Shimizu, K.~Takagi, M.~Tanimoto, T.H.~Tatsuishi and H.~Uchida,
  \emph{{Finite modular subgroups for fermion mass matrices and baryon/lepton
  number violation}},
  \href{https://doi.org/10.1016/j.physletb.2019.05.034}{\emph{Phys. Lett. B}
  {\bfseries 794} (2019) 114}
  [\href{https://arxiv.org/abs/1812.11072}{{\ttfamily 1812.11072}}].

\bibitem{Novichkov:2018yse}
P.P.~Novichkov, S.T.~Petcov and M.~Tanimoto, \emph{{Trimaximal Neutrino Mixing
  from Modular A4 Invariance with Residual Symmetries}},
  \href{https://doi.org/10.1016/j.physletb.2019.04.043}{\emph{Phys. Lett. B}
  {\bfseries 793} (2019) 247}
  [\href{https://arxiv.org/abs/1812.11289}{{\ttfamily 1812.11289}}].

\bibitem{Okada:2019uoy}
H.~Okada and M.~Tanimoto, \emph{{Towards unification of quark and lepton
  flavors in $A_4$ modular invariance}},
  \href{https://doi.org/10.1140/epjc/s10052-021-08845-y}{\emph{Eur. Phys. J. C}
  {\bfseries 81} (2021) 52} [\href{https://arxiv.org/abs/1905.13421}{{\ttfamily
  1905.13421}}].

\bibitem{Kobayashi:2019mna}
T.~Kobayashi, Y.~Shimizu, K.~Takagi, M.~Tanimoto and T.H.~Tatsuishi, \emph{{New
  $A_4$ lepton flavor model from $S_4$ modular symmetry}},
  \href{https://doi.org/10.1007/JHEP02(2020)097}{\emph{JHEP} {\bfseries 02}
  (2020) 097} [\href{https://arxiv.org/abs/1907.09141}{{\ttfamily
  1907.09141}}].

\bibitem{Ding:2019zxk}
G.-J.~Ding, S.F.~King and X.-G.~Liu, \emph{{Modular A$_{4}$ symmetry models of
  neutrinos and charged leptons}},
  \href{https://doi.org/10.1007/JHEP09(2019)074}{\emph{JHEP} {\bfseries 09}
  (2019) 074} [\href{https://arxiv.org/abs/1907.11714}{{\ttfamily
  1907.11714}}].

\bibitem{Okada:2019mjf}
H.~Okada and Y.~Orikasa, \emph{{A radiative seesaw model in modular $A_4$
  symmetry}},  \href{https://arxiv.org/abs/1907.13520}{{\ttfamily 1907.13520}}.

\bibitem{Kobayashi:2019xvz}
T.~Kobayashi, Y.~Shimizu, K.~Takagi, M.~Tanimoto and T.H.~Tatsuishi,
  \emph{{$A_4$ lepton flavor model and modulus stabilization from $S_4$ modular
  symmetry}}, \href{https://doi.org/10.1103/PhysRevD.100.115045}{\emph{Phys.
  Rev. D} {\bfseries 100} (2019) 115045}
  [\href{https://arxiv.org/abs/1909.05139}{{\ttfamily 1909.05139}}].

\bibitem{Chen:2019ewa}
M.-C.~Chen, S.~Ramos-S\'anchez and M.~Ratz, \emph{{A note on the predictions of
  models with modular flavor symmetries}},
  \href{https://doi.org/10.1016/j.physletb.2019.135153}{\emph{Phys. Lett. B}
  {\bfseries 801} (2020) 135153}
  [\href{https://arxiv.org/abs/1909.06910}{{\ttfamily 1909.06910}}].

\bibitem{Zhang:2019ngf}
D.~Zhang, \emph{{A modular $A_4$ symmetry realization of two-zero textures of
  the Majorana neutrino mass matrix}},
  \href{https://doi.org/10.1016/j.nuclphysb.2020.114935}{\emph{Nucl. Phys. B}
  {\bfseries 952} (2020) 114935}
  [\href{https://arxiv.org/abs/1910.07869}{{\ttfamily 1910.07869}}].

\bibitem{Nomura:2019xsb}
T.~Nomura, H.~Okada and S.~Patra, \emph{{An inverse seesaw model with $A_4$
  -modular symmetry}},
  \href{https://doi.org/10.1016/j.nuclphysb.2021.115395}{\emph{Nucl. Phys. B}
  {\bfseries 967} (2021) 115395}
  [\href{https://arxiv.org/abs/1912.00379}{{\ttfamily 1912.00379}}].

\bibitem{Kobayashi:2019gtp}
T.~Kobayashi, T.~Nomura and T.~Shimomura, \emph{{Type II seesaw models with
  modular $A_4$ symmetry}},
  \href{https://doi.org/10.1103/PhysRevD.102.035019}{\emph{Phys. Rev. D}
  {\bfseries 102} (2020) 035019}
  [\href{https://arxiv.org/abs/1912.00637}{{\ttfamily 1912.00637}}].

\bibitem{Wang:2019xbo}
X.~Wang, \emph{{Lepton flavor mixing and CP violation in the minimal
  type-(I+II) seesaw model with a modular $A_4$ symmetry}},
  \href{https://doi.org/10.1016/j.nuclphysb.2020.115105}{\emph{Nucl. Phys. B}
  {\bfseries 957} (2020) 115105}
  [\href{https://arxiv.org/abs/1912.13284}{{\ttfamily 1912.13284}}].

\bibitem{King:2020qaj}
S.J.D.~King and S.F.~King, \emph{{Fermion mass hierarchies from modular
  symmetry}}, \href{https://doi.org/10.1007/JHEP09(2020)043}{\emph{JHEP}
  {\bfseries 09} (2020) 043}
  [\href{https://arxiv.org/abs/2002.00969}{{\ttfamily 2002.00969}}].

\bibitem{Abbas:2020qzc}
M.~Abbas, \emph{{Fermion masses and mixing in modular A4 Symmetry}},
  \href{https://doi.org/10.1103/PhysRevD.103.056016}{\emph{Phys. Rev. D}
  {\bfseries 103} (2021) 056016}
  [\href{https://arxiv.org/abs/2002.01929}{{\ttfamily 2002.01929}}].

\bibitem{Okada:2020rjb}
H.~Okada and M.~Tanimoto, \emph{{Quark and lepton flavors with common modulus
  $\tau$ in $A_4$ modular symmetry}},
  \href{https://arxiv.org/abs/2005.00775}{{\ttfamily 2005.00775}}.

\bibitem{Nomura:2020opk}
T.~Nomura and H.~Okada, \emph{{A linear seesaw model with $A_4$-modular flavor
  and local $U(1)_{B-L}$ symmetries}},
  \href{https://arxiv.org/abs/2007.04801}{{\ttfamily 2007.04801}}.

\bibitem{Nomura:2020cog}
T.~Nomura and H.~Okada, \emph{{Modular $A_4$ symmetric inverse seesaw model
  with $SU(2)_L$ multiplet fields}},
  \href{https://arxiv.org/abs/2007.15459}{{\ttfamily 2007.15459}}.

\bibitem{Asaka:2020tmo}
T.~Asaka, Y.~Heo and T.~Yoshida, \emph{{Lepton flavor model with modular $A_4$
  symmetry in large volume limit}},
  \href{https://doi.org/10.1016/j.physletb.2020.135956}{\emph{Phys. Lett. B}
  {\bfseries 811} (2020) 135956}
  [\href{https://arxiv.org/abs/2009.12120}{{\ttfamily 2009.12120}}].

\bibitem{Nagao:2020snm}
K.I.~Nagao and H.~Okada, \emph{{Lepton sector in modular A4 and gauged U(1)R
  symmetry}},
  \href{https://doi.org/10.1016/j.nuclphysb.2022.115841}{\emph{Nucl. Phys. B}
  {\bfseries 980} (2022) 115841}
  [\href{https://arxiv.org/abs/2010.03348}{{\ttfamily 2010.03348}}].

\bibitem{Okada:2020brs}
H.~Okada and M.~Tanimoto, \emph{{Spontaneous CP violation by modulus $\tau$ in
  $A_4$ model of lepton flavors}},
  \href{https://doi.org/10.1007/JHEP03(2021)010}{\emph{JHEP} {\bfseries 03}
  (2021) 010} [\href{https://arxiv.org/abs/2012.01688}{{\ttfamily
  2012.01688}}].

\bibitem{Gehrlein:2020jnr}
J.~Gehrlein and M.~Spinrath, \emph{{Leptonic Sum Rules from Flavour Models with
  Modular Symmetries}},
  \href{https://doi.org/10.1007/JHEP03(2021)177}{\emph{JHEP} {\bfseries 03}
  (2021) 177} [\href{https://arxiv.org/abs/2012.04131}{{\ttfamily
  2012.04131}}].

\bibitem{Hutauruk:2020xtk}
P.T.P.~Hutauruk, D.W.~Kang, J.~Kim and H.~Okada, \emph{{Muon $g-2$ and neutrino
  mass explanations in a modular $A_4$ symmetry}},
  \href{https://arxiv.org/abs/2012.11156}{{\ttfamily 2012.11156}}.

\bibitem{Yao:2020qyy}
C.-Y.~Yao, J.-N.~Lu and G.-J.~Ding, \emph{{Modular Invariant $A_{4}$ Models for
  Quarks and Leptons with Generalized CP Symmetry}},
  \href{https://doi.org/10.1007/JHEP05(2021)102}{\emph{JHEP} {\bfseries 05}
  (2021) 102} [\href{https://arxiv.org/abs/2012.13390}{{\ttfamily
  2012.13390}}].

\bibitem{Feruglio:2021dte}
F.~Feruglio, V.~Gherardi, A.~Romanino and A.~Titov, \emph{{Modular invariant
  dynamics and fermion mass hierarchies around $\tau = i$}},
  \href{https://doi.org/10.1007/JHEP05(2021)242}{\emph{JHEP} {\bfseries 05}
  (2021) 242} [\href{https://arxiv.org/abs/2101.08718}{{\ttfamily
  2101.08718}}].

\bibitem{Kobayashi:2021jqu}
T.~Kobayashi, T.~Shimomura and M.~Tanimoto, \emph{{Soft supersymmetry breaking
  terms and lepton flavor violations in modular flavor models}},
  \href{https://doi.org/10.1016/j.physletb.2021.136452}{\emph{Phys. Lett. B}
  {\bfseries 819} (2021) 136452}
  [\href{https://arxiv.org/abs/2102.10425}{{\ttfamily 2102.10425}}].

\bibitem{Tanimoto:2021ehw}
M.~Tanimoto and K.~Yamamoto, \emph{{Electron EDM arising from modulus
  \ensuremath{\tau} in the supersymmetric modular invariant flavor models}},
  \href{https://doi.org/10.1007/JHEP10(2021)183}{\emph{JHEP} {\bfseries 10}
  (2021) 183} [\href{https://arxiv.org/abs/2106.10919}{{\ttfamily
  2106.10919}}].

\bibitem{Nomura:2021yjb}
T.~Nomura, H.~Okada and Y.~Orikasa, \emph{{Quark and lepton flavor model with
  leptoquarks in a modular $A_4$ symmetry}},
  \href{https://doi.org/10.1140/epjc/s10052-021-09667-8}{\emph{Eur. Phys. J. C}
  {\bfseries 81} (2021) 947}
  [\href{https://arxiv.org/abs/2106.12375}{{\ttfamily 2106.12375}}].

\bibitem{deMedeirosVarzielas:2021pug}
I.~de~Medeiros~Varzielas and J.a.~Louren\c{c}o, \emph{{Two A4 modular
  symmetries for Tri-Maximal 2 mixing}},
  \href{https://doi.org/10.1016/j.nuclphysb.2022.115793}{\emph{Nucl. Phys. B}
  {\bfseries 979} (2022) 115793}
  [\href{https://arxiv.org/abs/2107.04042}{{\ttfamily 2107.04042}}].

\bibitem{Chen:2021prl}
M.-C.~Chen, V.~Knapp-Perez, M.~Ramos-Hamud, S.~Ramos-Sanchez, M.~Ratz and
  S.~Shukla, \emph{{Quasi\textendash{}eclectic modular flavor symmetries}},
  \href{https://doi.org/10.1016/j.physletb.2021.136843}{\emph{Phys. Lett. B}
  {\bfseries 824} (2022) 136843}
  [\href{https://arxiv.org/abs/2108.02240}{{\ttfamily 2108.02240}}].

\bibitem{Okada:2021aoi}
H.~Okada and Y.-h.~Qi, \emph{{Zee-Babu model in modular $A_4$ symmetry}},
  \href{https://arxiv.org/abs/2109.13779}{{\ttfamily 2109.13779}}.

\bibitem{Nomura:2021pld}
T.~Nomura, H.~Okada and Y.-h.~Qi, \emph{{Zee model in a modular $A_4$
  symmetry}},  \href{https://arxiv.org/abs/2111.10944}{{\ttfamily 2111.10944}}.

\bibitem{Liu:2021gwa}
X.-G.~Liu and G.-J.~Ding, \emph{{Modular flavor symmetry and vector-valued
  modular forms}}, \href{https://doi.org/10.1007/JHEP03(2022)123}{\emph{JHEP}
  {\bfseries 03} (2022) 123}
  [\href{https://arxiv.org/abs/2112.14761}{{\ttfamily 2112.14761}}].

\bibitem{Kikuchi:2022txy}
S.~Kikuchi, T.~Kobayashi, H.~Otsuka, M.~Tanimoto, H.~Uchida and K.~Yamamoto,
  \emph{{4D modular flavor symmetric models inspired by higher dimensional
  theory}},  \href{https://arxiv.org/abs/2201.04505}{{\ttfamily 2201.04505}}.

\bibitem{Nomura:2022hxs}
T.~Nomura and H.~Okada, \emph{{A radiative seesaw model in a supersymmetric
  modular $A_4$ group}},  \href{https://arxiv.org/abs/2201.10244}{{\ttfamily
  2201.10244}}.

\bibitem{Otsuka:2022rak}
H.~Otsuka and H.~Okada, \emph{{Radiative neutrino masses from modular $A_4$
  symmetry and supersymmetry breaking}},
  \href{https://arxiv.org/abs/2202.10089}{{\ttfamily 2202.10089}}.

\bibitem{Kobayashi:2022jvy}
T.~Kobayashi, H.~Otsuka, M.~Tanimoto and K.~Yamamoto, \emph{{Lepton flavor
  violation, lepton $(g-2)_{\mu,\,e}$ and electron EDM in the modular
  symmetry}},  \href{https://arxiv.org/abs/2204.12325}{{\ttfamily 2204.12325}}.

\bibitem{Ahn:2022ufs}
Y.H.~Ahn, S.K.~Kang, R.~Ramos and M.~Tanimoto, \emph{{Confronting the
  prediction of leptonic Dirac CP-violating phase with experiments}},
  \href{https://arxiv.org/abs/2205.02796}{{\ttfamily 2205.02796}}.

\bibitem{Kashav:2022kpk}
M.~Kashav and S.~Verma, \emph{{On Minimal realization of Topological Lorentz
  Structures with one-loop Seesaw extensions in A$_4$ Modular Symmetry}},
  \href{https://arxiv.org/abs/2205.06545}{{\ttfamily 2205.06545}}.

\bibitem{Nomura:2022boj}
T.~Nomura, H.~Okada and Y.~Shoji, \emph{{$SU(4)_C \times SU(2)_L \times U(1)_R$
  models with modular $A_4$ symmetry}},
  \href{https://arxiv.org/abs/2206.04466}{{\ttfamily 2206.04466}}.

\bibitem{Ishiguro:2022pde}
K.~Ishiguro, H.~Okada and H.~Otsuka, \emph{{Residual flavor symmetry breaking
  in the landscape of modular flavor models}},
  \href{https://arxiv.org/abs/2206.04313}{{\ttfamily 2206.04313}}.

\bibitem{Gogoi:2022jwf}
J.~Gogoi, N.~Gautam and M.K.~Das, \emph{{Neutrino masses and mixing in Minimal
  Inverse Seesaw using $A_4$ modular symmetry}},
  \href{https://arxiv.org/abs/2207.10546}{{\ttfamily 2207.10546}}.

\bibitem{Novichkov:2018nkm}
P.P.~Novichkov, J.T.~Penedo, S.T.~Petcov and A.V.~Titov, \emph{{Modular A$_{5}$
  symmetry for flavour model building}},
  \href{https://doi.org/10.1007/JHEP04(2019)174}{\emph{JHEP} {\bfseries 04}
  (2019) 174} [\href{https://arxiv.org/abs/1812.02158}{{\ttfamily
  1812.02158}}].

\bibitem{Ding:2019xna}
G.-J.~Ding, S.F.~King and X.-G.~Liu, \emph{{Neutrino mass and mixing with $A_5$
  modular symmetry}},
  \href{https://doi.org/10.1103/PhysRevD.100.115005}{\emph{Phys. Rev. D}
  {\bfseries 100} (2019) 115005}
  [\href{https://arxiv.org/abs/1903.12588}{{\ttfamily 1903.12588}}].

\bibitem{deMedeirosVarzielas:2022ihu}
I.~de~Medeiros~Varzielas and J.a.~Louren\c{c}o, \emph{{Two A5 modular
  symmetries for Golden Ratio 2 mixing}},
  \href{https://arxiv.org/abs/2206.14869}{{\ttfamily 2206.14869}}.

\bibitem{Nilles:2020nnc}
H.P.~Nilles, S.~Ramos-S\'anchez and P.K.S.~Vaudrevange, \emph{{Eclectic Flavor
  Groups}}, \href{https://doi.org/10.1007/JHEP02(2020)045}{\emph{JHEP}
  {\bfseries 02} (2020) 045}
  [\href{https://arxiv.org/abs/2001.01736}{{\ttfamily 2001.01736}}].

\bibitem{Ding:2020msi}
G.-J.~Ding, S.F.~King, C.-C.~Li and Y.-L.~Zhou, \emph{{Modular Invariant Models
  of Leptons at Level 7}},
  \href{https://doi.org/10.1007/JHEP08(2020)164}{\emph{JHEP} {\bfseries 08}
  (2020) 164} [\href{https://arxiv.org/abs/2004.12662}{{\ttfamily
  2004.12662}}].

\bibitem{Li:2021buv}
C.-C.~Li, X.-G.~Liu and G.-J.~Ding, \emph{{Modular symmetry at level 6 and a
  new route towards finite modular groups}},
  \href{https://doi.org/10.1007/JHEP10(2021)238}{\emph{JHEP} {\bfseries 10}
  (2021) 238} [\href{https://arxiv.org/abs/2108.02181}{{\ttfamily
  2108.02181}}].

\bibitem{Ohki:2020bpo}
H.~Ohki, S.~Uemura and R.~Watanabe, \emph{{Modular flavor symmetry on a
  magnetized torus}},
  \href{https://doi.org/10.1103/PhysRevD.102.085008}{\emph{Phys. Rev. D}
  {\bfseries 102} (2020) 085008}
  [\href{https://arxiv.org/abs/2003.04174}{{\ttfamily 2003.04174}}].

\bibitem{Okada:2018yrn}
H.~Okada and M.~Tanimoto, \emph{{CP violation of quarks in $A_4$ modular
  invariance}},
  \href{https://doi.org/10.1016/j.physletb.2019.02.028}{\emph{Phys. Lett. B}
  {\bfseries 791} (2019) 54}
  [\href{https://arxiv.org/abs/1812.09677}{{\ttfamily 1812.09677}}].

\bibitem{Kuranaga:2021ujd}
H.~Kuranaga, H.~Ohki and S.~Uemura, \emph{{Modular origin of mass hierarchy:
  Froggatt-Nielsen like mechanism}},
  \href{https://doi.org/10.1007/JHEP07(2021)068}{\emph{JHEP} {\bfseries 07}
  (2021) 068} [\href{https://arxiv.org/abs/2105.06237}{{\ttfamily
  2105.06237}}].

\bibitem{Kikuchi:2022geu}
S.~Kikuchi, T.~Kobayashi, M.~Tanimoto and H.~Uchida, \emph{{Mass matrices with
  CP phase in modular flavor symmetry}},
  \href{https://arxiv.org/abs/2206.08538}{{\ttfamily 2206.08538}}.

\bibitem{Kikuchi:2022svo}
S.~Kikuchi, T.~Kobayashi, M.~Tanimoto and H.~Uchida, \emph{{Texture zeros of
  quark mass matrices at fixed point $\tau=\omega$ in modular flavor
  symmetry}},  \href{https://arxiv.org/abs/2207.04609}{{\ttfamily 2207.04609}}.

\bibitem{Kikuchi:2022pkd}
S.~Kikuchi, T.~Kobayashi, K.~Nasu, H.~Otsuka, S.~Takada and H.~Uchida,
  \emph{{Modular symmetry of soft supersymmetry breaking terms}},
  \href{https://arxiv.org/abs/2203.14667}{{\ttfamily 2203.14667}}.

\bibitem{Kobayashi:2019uyt}
T.~Kobayashi, Y.~Shimizu, K.~Takagi, M.~Tanimoto, T.H.~Tatsuishi and H.~Uchida,
  \emph{{$CP$ violation in modular invariant flavor models}},
  \href{https://doi.org/10.1103/PhysRevD.101.055046}{\emph{Phys. Rev. D}
  {\bfseries 101} (2020) 055046}
  [\href{https://arxiv.org/abs/1910.11553}{{\ttfamily 1910.11553}}].

\bibitem{deMedeirosVarzielas:2020kji}
I.~de~Medeiros~Varzielas, M.~Levy and Y.-L.~Zhou, \emph{{Symmetries and
  stabilisers in modular invariant flavour models}},
  \href{https://doi.org/10.1007/JHEP11(2020)085}{\emph{JHEP} {\bfseries 11}
  (2020) 085} [\href{https://arxiv.org/abs/2008.05329}{{\ttfamily
  2008.05329}}].

\bibitem{Ding:2020zxw}
G.-J.~Ding, F.~Feruglio and X.-G.~Liu, \emph{{Automorphic Forms and Fermion
  Masses}}, \href{https://doi.org/10.1007/JHEP01(2021)037}{\emph{JHEP}
  {\bfseries 01} (2021) 037}
  [\href{https://arxiv.org/abs/2010.07952}{{\ttfamily 2010.07952}}].

\bibitem{Ishiguro:2020tmo}
K.~Ishiguro, T.~Kobayashi and H.~Otsuka, \emph{{Landscape of Modular Symmetric
  Flavor Models}}, \href{https://doi.org/10.1007/JHEP03(2021)161}{\emph{JHEP}
  {\bfseries 03} (2021) 161}
  [\href{https://arxiv.org/abs/2011.09154}{{\ttfamily 2011.09154}}].

\bibitem{Novichkov:2022wvg}
P.P.~Novichkov, J.T.~Penedo and S.T.~Petcov, \emph{{Modular flavour symmetries
  and modulus stabilisation}},
  \href{https://doi.org/10.1007/JHEP03(2022)149}{\emph{JHEP} {\bfseries 03}
  (2022) 149} [\href{https://arxiv.org/abs/2201.02020}{{\ttfamily
  2201.02020}}].

\bibitem{deAnda:2018ecu}
F.J.~de~Anda, S.F.~King and E.~Perdomo, \emph{{$SU(5)$ grand unified theory
  with $A_4$ modular symmetry}},
  \href{https://doi.org/10.1103/PhysRevD.101.015028}{\emph{Phys. Rev. D}
  {\bfseries 101} (2020) 015028}
  [\href{https://arxiv.org/abs/1812.05620}{{\ttfamily 1812.05620}}].

\bibitem{Kobayashi:2019rzp}
T.~Kobayashi, Y.~Shimizu, K.~Takagi, M.~Tanimoto and T.H.~Tatsuishi,
  \emph{{Modular $S_3$-invariant flavor model in SU(5) grand unified theory}},
  \href{https://doi.org/10.1093/ptep/ptaa055}{\emph{PTEP} {\bfseries 2020}
  (2020) 053B05} [\href{https://arxiv.org/abs/1906.10341}{{\ttfamily
  1906.10341}}].

\bibitem{Du:2020ylx}
X.~Du and F.~Wang, \emph{{SUSY breaking constraints on modular flavor $S_{3}$
  invariant SU(5) GUT model}},
  \href{https://doi.org/10.1007/JHEP02(2021)221}{\emph{JHEP} {\bfseries 02}
  (2021) 221} [\href{https://arxiv.org/abs/2012.01397}{{\ttfamily
  2012.01397}}].

\bibitem{Zhao:2021jxg}
Y.~Zhao and H.-H.~Zhang, \emph{{Adjoint SU(5) GUT model with modular $S_{4}$
  symmetry}}, \href{https://doi.org/10.1007/JHEP03(2021)002}{\emph{JHEP}
  {\bfseries 03} (2021) 002}
  [\href{https://arxiv.org/abs/2101.02266}{{\ttfamily 2101.02266}}].

\bibitem{Chen:2021zty}
P.~Chen, G.-J.~Ding and S.F.~King, \emph{{SU(5) GUTs with A$_{4}$ modular
  symmetry}}, \href{https://doi.org/10.1007/JHEP04(2021)239}{\emph{JHEP}
  {\bfseries 04} (2021) 239}
  [\href{https://arxiv.org/abs/2101.12724}{{\ttfamily 2101.12724}}].

\bibitem{King:2021fhl}
S.F.~King and Y.-L.~Zhou, \emph{{Twin modular S$_{4}$ with SU(5) GUT}},
  \href{https://doi.org/10.1007/JHEP04(2021)291}{\emph{JHEP} {\bfseries 04}
  (2021) 291} [\href{https://arxiv.org/abs/2103.02633}{{\ttfamily
  2103.02633}}].

\bibitem{Ding:2021zbg}
G.-J.~Ding, S.F.~King and C.-Y.~Yao, \emph{{Modular $S_4\times SU(5)$ GUT}},
  \href{https://doi.org/10.1103/PhysRevD.104.055034}{\emph{Phys. Rev. D}
  {\bfseries 104} (2021) 055034}
  [\href{https://arxiv.org/abs/2103.16311}{{\ttfamily 2103.16311}}].

\bibitem{Ding:2021eva}
G.-J.~Ding, S.F.~King and J.-N.~Lu, \emph{{SO(10) models with A$_{4}$ modular
  symmetry}}, \href{https://doi.org/10.1007/JHEP11(2021)007}{\emph{JHEP}
  {\bfseries 11} (2021) 007}
  [\href{https://arxiv.org/abs/2108.09655}{{\ttfamily 2108.09655}}].

\bibitem{Kobayashi:2022sov}
T.~Kobayashi, S.~Nishimura, H.~Otsuka, M.~Tanimoto and K.~Yamamoto,
  \emph{{Generalized Matter Parities from Finite Modular Symmetries}},
  \href{https://arxiv.org/abs/2207.14014}{{\ttfamily 2207.14014}}.

\bibitem{Nomura:2019jxj}
T.~Nomura and H.~Okada, \emph{{A modular $A_4$ symmetric model of dark matter
  and neutrino}},
  \href{https://doi.org/10.1016/j.physletb.2019.134799}{\emph{Phys. Lett. B}
  {\bfseries 797} (2019) 134799}
  [\href{https://arxiv.org/abs/1904.03937}{{\ttfamily 1904.03937}}].

\bibitem{Nomura:2019yft}
T.~Nomura and H.~Okada, \emph{{A two loop induced neutrino mass model with
  modular $A_4$ symmetry}},
  \href{https://doi.org/10.1016/j.nuclphysb.2021.115372}{\emph{Nucl. Phys. B}
  {\bfseries 966} (2021) 115372}
  [\href{https://arxiv.org/abs/1906.03927}{{\ttfamily 1906.03927}}].

\bibitem{Okada:2019xqk}
H.~Okada and Y.~Orikasa, \emph{{Modular $S_3$ symmetric radiative seesaw
  model}}, \href{https://doi.org/10.1103/PhysRevD.100.115037}{\emph{Phys. Rev.
  D} {\bfseries 100} (2019) 115037}
  [\href{https://arxiv.org/abs/1907.04716}{{\ttfamily 1907.04716}}].

\bibitem{Nomura:2019lnr}
T.~Nomura, H.~Okada and O.~Popov, \emph{{A modular $A_4$ symmetric scotogenic
  model}}, \href{https://doi.org/10.1016/j.physletb.2020.135294}{\emph{Phys.
  Lett. B} {\bfseries 803} (2020) 135294}
  [\href{https://arxiv.org/abs/1908.07457}{{\ttfamily 1908.07457}}].

\bibitem{Okada:2019lzv}
H.~Okada and Y.~Orikasa, \emph{{Neutrino mass model with a modular $S_4$
  symmetry}},  \href{https://arxiv.org/abs/1908.08409}{{\ttfamily 1908.08409}}.

\bibitem{Okada:2020dmb}
H.~Okada and Y.~Shoji, \emph{{A radiative seesaw model with three Higgs
  doublets in modular $A_4$ symmetry}},
  \href{https://doi.org/10.1016/j.nuclphysb.2020.115216}{\emph{Nucl. Phys. B}
  {\bfseries 961} (2020) 115216}
  [\href{https://arxiv.org/abs/2003.13219}{{\ttfamily 2003.13219}}].

\bibitem{Nagao:2021rio}
K.I.~Nagao and H.~Okada, \emph{{Modular A4 symmetry and light dark matter with
  gauged U(1)B\ensuremath{-}L}},
  \href{https://doi.org/10.1016/j.dark.2022.101039}{\emph{Phys. Dark Univ.}
  {\bfseries 36} (2022) 101039}
  [\href{https://arxiv.org/abs/2108.09984}{{\ttfamily 2108.09984}}].

\bibitem{Kobayashi:2021ajl}
T.~Kobayashi, H.~Okada and Y.~Orikasa, \emph{{Dark matter stability at fixed
  points in a modular $A_4$ symmetry}},
  \href{https://arxiv.org/abs/2111.05674}{{\ttfamily 2111.05674}}.

\bibitem{Asaka:2019vev}
T.~Asaka, Y.~Heo, T.H.~Tatsuishi and T.~Yoshida, \emph{{Modular $A_4$
  invariance and leptogenesis}},
  \href{https://doi.org/10.1007/JHEP01(2020)144}{\emph{JHEP} {\bfseries 01}
  (2020) 144} [\href{https://arxiv.org/abs/1909.06520}{{\ttfamily
  1909.06520}}].

\bibitem{Behera:2020sfe}
M.K.~Behera, S.~Mishra, S.~Singirala and R.~Mohanta, \emph{{Implications of A4
  modular symmetry on neutrino mass, mixing and leptogenesis with linear
  seesaw}}, \href{https://doi.org/10.1016/j.dark.2022.101027}{\emph{Phys. Dark
  Univ.} {\bfseries 36} (2022) 101027}
  [\href{https://arxiv.org/abs/2007.00545}{{\ttfamily 2007.00545}}].

\bibitem{Mishra:2020gxg}
S.~Mishra, \emph{{Neutrino mixing and Leptogenesis with modular $S_3$ symmetry
  in the framework of type III seesaw}},
  \href{https://arxiv.org/abs/2008.02095}{{\ttfamily 2008.02095}}.

\bibitem{Kashav:2021zir}
M.~Kashav and S.~Verma, \emph{{Broken scaling neutrino mass matrix and
  leptogenesis based on A$_{4}$ modular invariance}},
  \href{https://doi.org/10.1007/JHEP09(2021)100}{\emph{JHEP} {\bfseries 09}
  (2021) 100} [\href{https://arxiv.org/abs/2103.07207}{{\ttfamily
  2103.07207}}].

\bibitem{Okada:2021qdf}
H.~Okada, Y.~Shimizu, M.~Tanimoto and T.~Yoshida, \emph{{Modulus
  \ensuremath{\tau} linking leptonic CP violation to baryon asymmetry in
  A$_{4}$ modular invariant flavor model}},
  \href{https://doi.org/10.1007/JHEP07(2021)184}{\emph{JHEP} {\bfseries 07}
  (2021) 184} [\href{https://arxiv.org/abs/2105.14292}{{\ttfamily
  2105.14292}}].

\bibitem{Dasgupta:2021ggp}
A.~Dasgupta, T.~Nomura, H.~Okada, O.~Popov and M.~Tanimoto, \emph{{Dirac
  Radiative Neutrino Mass with Modular Symmetry and Leptogenesis}},
  \href{https://arxiv.org/abs/2111.06898}{{\ttfamily 2111.06898}}.

\bibitem{Behera:2022wco}
M.K.~Behera and R.~Mohanta, \emph{{Linear Seesaw in A5' Modular Symmetry With
  Leptogenesis}}, \href{https://doi.org/10.3389/fphy.2022.854595}{\emph{Front.
  in Phys.} {\bfseries 10} (2022) 854595}
  [\href{https://arxiv.org/abs/2201.10429}{{\ttfamily 2201.10429}}].

\bibitem{Kang:2022psa}
D.W.~Kang, J.~Kim, T.~Nomura and H.~Okada, \emph{{Natural mass hierarchy among
  three heavy Majorana neutrinos for resonant leptogenesis under modular $A_4$
  symmetry}},  \href{https://arxiv.org/abs/2205.08269}{{\ttfamily 2205.08269}}.

\bibitem{Ding:2022bzs}
G.-J.~Ding, S.F.~King, J.-N.~Lu and B.-Y.~Qu, \emph{{Leptogenesis in $SO(10)$
  Models with $A_4$ Modular Symmetry}},
  \href{https://arxiv.org/abs/2206.14675}{{\ttfamily 2206.14675}}.

\bibitem{Starobinsky:1980te}
A.A.~Starobinsky, \emph{{A New Type of Isotropic Cosmological Models Without
  Singularity}}, \href{https://doi.org/10.1016/0370-2693(80)90670-X}{\emph{Adv.
  Ser. Astrophys. Cosmol.} {\bfseries 3} (1987) 130}.

\bibitem{Mukhanov:1981xt}
V.F.~Mukhanov and G.V.~Chibisov, \emph{{Quantum Fluctuations and a Nonsingular
  Universe}}, {\emph{JETP Lett.} {\bfseries 33} (1981) 532}.

\bibitem{Buchmuller:2013zfa}
W.~Buchmuller, V.~Domcke and K.~Kamada, \emph{{The Starobinsky Model from
  Superconformal D-Term Inflation}},
  \href{https://doi.org/10.1016/j.physletb.2013.08.042}{\emph{Phys. Lett. B}
  {\bfseries 726} (2013) 467}
  [\href{https://arxiv.org/abs/1306.3471}{{\ttfamily 1306.3471}}].

\bibitem{Buchmuller:2014rfa}
W.~Buchmuller, V.~Domcke and K.~Schmitz, \emph{{The Chaotic Regime of D-Term
  Inflation}}, \href{https://doi.org/10.1088/1475-7516/2014/11/006}{\emph{JCAP}
  {\bfseries 11} (2014) 006} [\href{https://arxiv.org/abs/1406.6300}{{\ttfamily
  1406.6300}}].

\bibitem{Buchmuller:2014dda}
W.~Buchmuller and K.~Ishiwata, \emph{{Grand Unification and Subcritical Hybrid
  Inflation}}, \href{https://doi.org/10.1103/PhysRevD.91.081302}{\emph{Phys.
  Rev. D} {\bfseries 91} (2015) 081302}
  [\href{https://arxiv.org/abs/1412.3764}{{\ttfamily 1412.3764}}].

\bibitem{Kallosh:2013yoa}
R.~Kallosh, A.~Linde and D.~Roest, \emph{{Superconformal Inflationary
  $\alpha$-Attractors}},
  \href{https://doi.org/10.1007/JHEP11(2013)198}{\emph{JHEP} {\bfseries 11}
  (2013) 198} [\href{https://arxiv.org/abs/1311.0472}{{\ttfamily 1311.0472}}].

\bibitem{Ishiwata:2018dxg}
K.~Ishiwata, \emph{{Superconformal Subcritical Hybrid Inflation}},
  \href{https://doi.org/10.1016/j.physletb.2018.05.047}{\emph{Phys. Lett. B}
  {\bfseries 782} (2018) 367}
  [\href{https://arxiv.org/abs/1803.08274}{{\ttfamily 1803.08274}}].

\bibitem{Gunji:2021zit}
Y.~Gunji and K.~Ishiwata, \emph{{Subcritical hybrid inflation in a generalized
  superconformal model}},
  \href{https://doi.org/10.1103/PhysRevD.104.123545}{\emph{Phys. Rev. D}
  {\bfseries 104} (2021) 123545}
  [\href{https://arxiv.org/abs/2104.02248}{{\ttfamily 2104.02248}}].

\bibitem{deAdelhartToorop:2011re}
R.~de~Adelhart~Toorop, F.~Feruglio and C.~Hagedorn, \emph{{Finite Modular
  Groups and Lepton Mixing}},
  \href{https://doi.org/10.1016/j.nuclphysb.2012.01.017}{\emph{Nucl. Phys. B}
  {\bfseries 858} (2012) 437}
  [\href{https://arxiv.org/abs/1112.1340}{{\ttfamily 1112.1340}}].

\bibitem{Gunji:2019wtk}
Y.~Gunji and K.~Ishiwata, \emph{{Leptogenesis after superconformal subcritical
  hybrid inflation}},
  \href{https://doi.org/10.1007/JHEP09(2019)065}{\emph{JHEP} {\bfseries 09}
  (2019) 065} [\href{https://arxiv.org/abs/1906.04530}{{\ttfamily
  1906.04530}}].

\bibitem{Ferrara:2010in}
S.~Ferrara, R.~Kallosh, A.~Linde, A.~Marrani and A.~Van~Proeyen,
  \emph{{Superconformal Symmetry, NMSSM, and Inflation}},
  \href{https://doi.org/10.1103/PhysRevD.83.025008}{\emph{Phys. Rev. D}
  {\bfseries 83} (2011) 025008}
  [\href{https://arxiv.org/abs/1008.2942}{{\ttfamily 1008.2942}}].

\bibitem{Minkowski:1977sc}
P.~Minkowski, \emph{{$\mu \to e\gamma$ at a Rate of One Out of $10^{9}$ Muon
  Decays?}}, \href{https://doi.org/10.1016/0370-2693(77)90435-X}{\emph{Phys.
  Lett. B} {\bfseries 67} (1977) 421}.

\bibitem{Yanagida:1979as}
T.~Yanagida, \emph{{Horizontal gauge symmetry and masses of neutrinos}},
  {\emph{Conf. Proc. C} {\bfseries 7902131} (1979) 95}.

\bibitem{Yanagida:1980xy}
T.~Yanagida, \emph{{Horizontal Symmetry and Masses of Neutrinos}},
  \href{https://doi.org/10.1143/PTP.64.1103}{\emph{Prog. Theor. Phys.}
  {\bfseries 64} (1980) 1103}.

\bibitem{Gell-Mann:1979vob}
M.~Gell-Mann, P.~Ramond and R.~Slansky, \emph{{Complex Spinors and Unified
  Theories}}, {\emph{Conf. Proc. C} {\bfseries 790927} (1979) 315}
  [\href{https://arxiv.org/abs/1306.4669}{{\ttfamily 1306.4669}}].

\bibitem{Ramond:1979py}
P.~Ramond, \emph{{The Family Group in Grand Unified Theories}},  in
  \emph{{International Symposium on Fundamentals of Quantum Theory and Quantum
  Field Theory}}, 2, 1979
  [\href{https://arxiv.org/abs/hep-ph/9809459}{{\ttfamily hep-ph/9809459}}].

\bibitem{Glashow:1979nm}
S.L.~Glashow, \emph{{The Future of Elementary Particle Physics}},
  \href{https://doi.org/10.1007/978-1-4684-7197-7_15}{\emph{NATO Sci. Ser. B}
  {\bfseries 61} (1980) 687}.

\bibitem{Aghanim:2018eyx}
{\scshape Planck} collaboration, \emph{{Planck 2018 results. VI. Cosmological
  parameters}},
  \href{https://doi.org/10.1051/0004-6361/201833910}{\emph{Astron. Astrophys.}
  {\bfseries 641} (2020) A6}
  [\href{https://arxiv.org/abs/1807.06209}{{\ttfamily 1807.06209}}].

\bibitem{Vagnozzi:2017ovm}
S.~Vagnozzi, E.~Giusarma, O.~Mena, K.~Freese, M.~Gerbino, S.~Ho et~al.,
  \emph{{Unveiling $\nu$ secrets with cosmological data: neutrino masses and
  mass hierarchy}},
  \href{https://doi.org/10.1103/PhysRevD.96.123503}{\emph{Phys. Rev. D}
  {\bfseries 96} (2017) 123503}
  [\href{https://arxiv.org/abs/1701.08172}{{\ttfamily 1701.08172}}].

\bibitem{Esteban:2020cvm}
I.~Esteban, M.C.~Gonzalez-Garcia, M.~Maltoni, T.~Schwetz and A.~Zhou,
  \emph{{The fate of hints: updated global analysis of three-flavor neutrino
  oscillations}}, \href{https://doi.org/10.1007/JHEP09(2020)178}{\emph{JHEP}
  {\bfseries 09} (2020) 178}
  [\href{https://arxiv.org/abs/2007.14792}{{\ttfamily 2007.14792}}].

\bibitem{Jarlskog:1985ht}
C.~Jarlskog, \emph{{Commutator of the Quark Mass Matrices in the Standard
  Electroweak Model and a Measure of Maximal $CP$~Nonconservation}},
  \href{https://doi.org/10.1103/PhysRevLett.55.1039}{\emph{Phys. Rev. Lett.}
  {\bfseries 55} (1985) 1039}.

\bibitem{Bilenky:2001rz}
S.M.~Bilenky, S.~Pascoli and S.T.~Petcov, \emph{{Majorana neutrinos, neutrino
  mass spectrum, CP violation and neutrinoless double beta decay. 1. The Three
  neutrino mixing case}},
  \href{https://doi.org/10.1103/PhysRevD.64.053010}{\emph{Phys. Rev. D}
  {\bfseries 64} (2001) 053010}
  [\href{https://arxiv.org/abs/hep-ph/0102265}{{\ttfamily hep-ph/0102265}}].

\bibitem{Nieves:2001fc}
J.F.~Nieves and P.B.~Pal, \emph{{Rephasing invariant CP violating parameters
  with Majorana neutrinos}},
  \href{https://doi.org/10.1103/PhysRevD.64.076005}{\emph{Phys. Rev. D}
  {\bfseries 64} (2001) 076005}
  [\href{https://arxiv.org/abs/hep-ph/0105305}{{\ttfamily hep-ph/0105305}}].

\bibitem{Aguilar-Saavedra:2000jom}
J.A.~Aguilar-Saavedra and G.C.~Branco, \emph{{Unitarity triangles and
  geometrical description of CP violation with Majorana neutrinos}},
  \href{https://doi.org/10.1103/PhysRevD.62.096009}{\emph{Phys. Rev. D}
  {\bfseries 62} (2000) 096009}
  [\href{https://arxiv.org/abs/hep-ph/0007025}{{\ttfamily hep-ph/0007025}}].

\bibitem{Girardi:2016zwz}
I.~Girardi, S.T.~Petcov and A.V.~Titov, \emph{{Predictions for the Majorana CP
  Violation Phases in the Neutrino Mixing Matrix and Neutrinoless Double Beta
  Decay}}, \href{https://doi.org/10.1016/j.nuclphysb.2016.08.019}{\emph{Nucl.
  Phys. B} {\bfseries 911} (2016) 754}
  [\href{https://arxiv.org/abs/1605.04172}{{\ttfamily 1605.04172}}].

\bibitem{Giudice:2004tc}
G.F.~Giudice and A.~Romanino, \emph{{Split supersymmetry}},
  \href{https://doi.org/10.1016/j.nuclphysb.2004.08.001}{\emph{Nucl. Phys. B}
  {\bfseries 699} (2004) 65}
  [\href{https://arxiv.org/abs/hep-ph/0406088}{{\ttfamily hep-ph/0406088}}].

\bibitem{Giudice:2011cg}
G.F.~Giudice and A.~Strumia, \emph{{Probing High-Scale and Split Supersymmetry
  with Higgs Mass Measurements}},
  \href{https://doi.org/10.1016/j.nuclphysb.2012.01.001}{\emph{Nucl. Phys. B}
  {\bfseries 858} (2012) 63} [\href{https://arxiv.org/abs/1108.6077}{{\ttfamily
  1108.6077}}].

\bibitem{Pas:2015eia}
H.~P\"as and W.~Rodejohann, \emph{{Neutrinoless Double Beta Decay}},
  \href{https://doi.org/10.1088/1367-2630/17/11/115010}{\emph{New J. Phys.}
  {\bfseries 17} (2015) 115010}
  [\href{https://arxiv.org/abs/1507.00170}{{\ttfamily 1507.00170}}].

\bibitem{KamLAND-Zen:2022tow}
{\scshape KamLAND-Zen} collaboration, \emph{{First Search for the Majorana
  Nature of Neutrinos in the Inverted Mass Ordering Region with KamLAND-Zen}},
  \href{https://arxiv.org/abs/2203.02139}{{\ttfamily 2203.02139}}.

\bibitem{Asaka:2018hyk}
T.~Asaka and T.~Yoshida, \emph{{Resonant leptogenesis at TeV-scale and
  neutrinoless double beta decay}},
  \href{https://doi.org/10.1007/JHEP09(2019)089}{\emph{JHEP} {\bfseries 09}
  (2019) 089} [\href{https://arxiv.org/abs/1812.11323}{{\ttfamily
  1812.11323}}].

\bibitem{Fukugita:1986hr}
M.~Fukugita and T.~Yanagida, \emph{{Baryogenesis Without Grand Unification}},
  \href{https://doi.org/10.1016/0370-2693(86)91126-3}{\emph{Phys. Lett. B}
  {\bfseries 174} (1986) 45}.

\bibitem{Asaka:2001ez}
T.~Asaka, W.~Buchmuller and L.~Covi, \emph{{False vacuum decay after
  inflation}}, \href{https://doi.org/10.1016/S0370-2693(01)00623-2}{\emph{Phys.
  Lett. B} {\bfseries 510} (2001) 271}
  [\href{https://arxiv.org/abs/hep-ph/0104037}{{\ttfamily hep-ph/0104037}}].

\bibitem{Nakayama:2016gvg}
K.~Nakayama, F.~Takahashi and T.T.~Yanagida, \emph{{Viable Chaotic Inflation as
  a Source of Neutrino Masses and Leptogenesis}},
  \href{https://doi.org/10.1016/j.physletb.2016.03.051}{\emph{Phys. Lett. B}
  {\bfseries 757} (2016) 32}
  [\href{https://arxiv.org/abs/1601.00192}{{\ttfamily 1601.00192}}].

\bibitem{Bjorkeroth:2016qsk}
F.~Bj\"orkeroth, S.F.~King, K.~Schmitz and T.T.~Yanagida, \emph{{Leptogenesis
  after Chaotic Sneutrino Inflation and the Supersymmetry Breaking Scale}},
  \href{https://doi.org/10.1016/j.nuclphysb.2017.01.017}{\emph{Nucl. Phys. B}
  {\bfseries 916} (2017) 688}
  [\href{https://arxiv.org/abs/1608.04911}{{\ttfamily 1608.04911}}].

\bibitem{Murayama:1992ua}
H.~Murayama, H.~Suzuki, T.~Yanagida and J.~Yokoyama, \emph{{Chaotic inflation
  and baryogenesis by right-handed sneutrinos}},
  \href{https://doi.org/10.1103/PhysRevLett.70.1912}{\emph{Phys. Rev. Lett.}
  {\bfseries 70} (1993) 1912}.

\bibitem{Murayama:1993xu}
H.~Murayama, H.~Suzuki, T.~Yanagida and J.~Yokoyama, \emph{{Chaotic inflation
  and baryogenesis in supergravity}},
  \href{https://doi.org/10.1103/PhysRevD.50.R2356}{\emph{Phys. Rev. D}
  {\bfseries 50} (1994) R2356}
  [\href{https://arxiv.org/abs/hep-ph/9311326}{{\ttfamily hep-ph/9311326}}].

\bibitem{Murayama:1993em}
H.~Murayama and T.~Yanagida, \emph{{Leptogenesis in supersymmetric standard
  model with right-handed neutrino}},
  \href{https://doi.org/10.1016/0370-2693(94)91164-9}{\emph{Phys. Lett. B}
  {\bfseries 322} (1994) 349}
  [\href{https://arxiv.org/abs/hep-ph/9310297}{{\ttfamily hep-ph/9310297}}].

\bibitem{Hamaguchi:2001gw}
K.~Hamaguchi, H.~Murayama and T.~Yanagida, \emph{{Leptogenesis from N dominated
  early universe}},
  \href{https://doi.org/10.1103/PhysRevD.65.043512}{\emph{Phys. Rev. D}
  {\bfseries 65} (2002) 043512}
  [\href{https://arxiv.org/abs/hep-ph/0109030}{{\ttfamily hep-ph/0109030}}].

\bibitem{Ellis:2003sq}
J.R.~Ellis, M.~Raidal and T.~Yanagida, \emph{{Sneutrino inflation in the light
  of WMAP: Reheating, leptogenesis and flavor violating lepton decays}},
  \href{https://doi.org/10.1016/j.physletb.2003.11.029}{\emph{Phys. Lett. B}
  {\bfseries 581} (2004) 9}
  [\href{https://arxiv.org/abs/hep-ph/0303242}{{\ttfamily hep-ph/0303242}}].

\bibitem{Antusch:2004hd}
S.~Antusch, M.~Bastero-Gil, S.F.~King and Q.~Shafi, \emph{{Sneutrino hybrid
  inflation in supergravity}},
  \href{https://doi.org/10.1103/PhysRevD.71.083519}{\emph{Phys. Rev. D}
  {\bfseries 71} (2005) 083519}
  [\href{https://arxiv.org/abs/hep-ph/0411298}{{\ttfamily hep-ph/0411298}}].

\bibitem{Antusch:2009ty}
S.~Antusch, M.~Bastero-Gil, K.~Dutta, S.F.~King and P.M.~Kostka, \emph{{Chaotic
  Inflation in Supergravity with Heisenberg Symmetry}},
  \href{https://doi.org/10.1016/j.physletb.2009.08.022}{\emph{Phys. Lett. B}
  {\bfseries 679} (2009) 428}
  [\href{https://arxiv.org/abs/0905.0905}{{\ttfamily 0905.0905}}].

\bibitem{Kadota:2005mt}
K.~Kadota and J.~Yokoyama, \emph{{D-term inflation and leptogenesis by
  right-handed sneutrino}},
  \href{https://doi.org/10.1103/PhysRevD.73.043507}{\emph{Phys. Rev. D}
  {\bfseries 73} (2006) 043507}
  [\href{https://arxiv.org/abs/hep-ph/0512221}{{\ttfamily hep-ph/0512221}}].

\bibitem{Nakayama:2013nya}
K.~Nakayama, F.~Takahashi and T.T.~Yanagida, \emph{{Chaotic Inflation with
  Right-handed Sneutrinos after Planck}},
  \href{https://doi.org/10.1016/j.physletb.2014.01.022}{\emph{Phys. Lett. B}
  {\bfseries 730} (2014) 24} [\href{https://arxiv.org/abs/1311.4253}{{\ttfamily
  1311.4253}}].

\end{thebibliography}\endgroup
\bibliographystyle{JHEP}
\end{document}